\documentclass[twocolumn]{aastex61}
\usepackage[utf8]{inputenc}
\usepackage{tikz}
\usepackage{amsmath}
\usepackage{amssymb}

\newcommand\figurewidth{\columnwidth}
\definecolor{darkred}{RGB}{180,20,20}
\newcommand\changed[1]{{#1}}

\shorttitle{L-shell photoionization of Fe$^+$}
\shortauthors{Schippers et al.}
\begin{document}

\title{Near L-Edge Single and Multiple Photoionization of Singly Charged Iron Ions}

\correspondingauthor{Stefan Schippers}
\email{stefan.schippers@physik.uni-giessen.de}

\author[0000-0002-6166-7138]{Stefan Schippers}
\affiliation{I. Physikalisches Institut, Justus-Liebig-Universit\"{a}t Gie{\ss}en, Heinrich-Buff-Ring 16, 35392 Giessen, Germany}

\author[0000-0002-1228-5029]{Michael Martins}
\affiliation{Institut f\"{u}r Experimentalphysik, Universit\"{a}t Hamburg, Luruper Chaussee 149, 22761 Hamburg, Germany}

\author[0000-0001-5100-4229]{Randolf Beerwerth}
\affiliation{Helmholtz-Institut Jena, Fr\"obelstieg 3, D-07743 Jena, Germany}
\affiliation{Theoretisch-Physikalisches Institut, Friedrich-Schiller-Universit\"at Jena, D-07743 Jena, Germany}

\author[0000-0003-3985-2051]{Sadia Bari}
\affiliation{FS-SCS, DESY, Notkestra{\ss}e 85, 22607 Hamburg, Germany}

\author[0000-0001-8809-1696]{Kristof Holste}
\affiliation{I. Physikalisches Institut, Justus-Liebig-Universit\"{a}t Gie{\ss}en, Heinrich-Buff-Ring 16, 35392 Giessen, Germany}

\author{Kaja Schubert}
\affiliation{Institut f\"{u}r Experimentalphysik, Universit\"{a}t Hamburg, Luruper Chaussee 149, 22761 Hamburg, Germany}
\affiliation{FS-SCS, DESY, Notkestra{\ss}e 85, 22607 Hamburg, Germany}

\author[0000-0003-1154-0750]{Jens~Viefhaus}
\affiliation{FS-PE, DESY, Notkestra{\ss}e 85, 22607 Hamburg, Germany}

\author[0000-0002-1111-6610]{Daniel Wolf Savin}
\affiliation{Columbia Astrophysics Laboratory, Columbia University, 550 West 120th Street, New York, NY 10027, USA}

\author[0000-0003-3101-2824]{Stephan Fritzsche}
\affiliation{Helmholtz-Institut Jena, Fr\"obelstieg 3, D-07743 Jena, Germany}
\affiliation{Theoretisch-Physikalisches Institut, Friedrich-Schiller-Universit\"at Jena, D-07743 Jena, Germany}

\author[0000-0002-0030-6929]{Alfred M\"{u}ller}
\affiliation{Institut f\"{u}r Atom- und Molek\"{u}lphysik, Justus-Liebig-Universit\"{a}t Gie{\ss}en, Leihgesterner Weg 217, 35392 Giessen, Germany}



\begin{abstract}
Absolute cross sections for $m$-fold photoionization ($m=1,\ldots,6$) of Fe$^{+}$ by a single photon were measured employing the photon-ion merged-beams setup PIPE at the PETRA\,III synchrotron light source, operated by DESY in Hamburg, Germany. Photon energies were in the range 680--920 eV which covers the photoionization resonances associated with $2p$ and $2s$ excitation to higher atomic shells as well as the thresholds for $2p$ and $2s$ ionization. The corresponding resonance positions were measured with an uncertainty of $\pm 0.2$ eV. The cross section for Fe$^+$ photoabsorption is derived as the sum of the individually measured cross-sections for $m$-fold ionization. Calculations of the Fe$^+$ absorption cross sections have been carried out using two different theoretical approaches, Hartree-Fock including relativistic extensions and fully relativistic Multi-Configuration Dirac Fock. Apart from overall energy shifts of up to about 3~eV, the theoretical cross sections are in good agreement with each other and with the experimental results. In addition, the complex deexcitation cascades after the creation of inner-shell holes in the Fe$^+$ ion have been tracked on the atomic fine-structure level. The corresponding theoretical results for the product charge-state distributions are in much better agreement with the experimental data than  previously published configuration-average results. The present experimental and theoretical results are valuable for opacity calculations and are expected to pave the way to a more accurate determination of the iron abundance in the interstellar medium.
\end{abstract}




\section{Introduction}

Astrophysical observations of ultraviolet (UV) photon emission from neutral and ionized Fe atoms are used to trace the chemical evolution of the Cosmos and to constrain the mass distribution of the progenitor stars, which synthesized the Fe \citep{Jenkins2009}. However, a large, but largely unknown, fraction of the interstellar Fe is locked up in dust grains, which are not detectable in the UV. In order to better quantify the interstellar Fe abundance, astronomers have recently initiated a series of X-ray observations \citep{Juett2006}. X-ray observations of Fe $L$-shell features facilitate the detection of Fe both in the gas phase and in the solid phase (i.e., dust grains). The fraction of Fe that is locked into grains is an important parameter to trace the chemical history of the Universe and any evolution in the stellar mass distribution.

The spectral features from molecules and solids are expected to differ from those of atoms, because the valence-shell electronic levels of molecules and solids are more fully occupied than those for atoms \citep{Gatuzz2016}. Hence, an accurate modeling of the atomic components is critical for inferring the composition of any molecular or solid phase Fe in the interstellar medium (ISM). Unfortunately crucial atomic data are lacking in current X-ray photoabsorption models for the ISM \citep{Wilms2000,Gatuzz2015}.

Near-neutral charge states of Fe are expected to be the dominant gas-phase form of Fe for most regions in the ISM, as has been shown through UV observations \citep{Jensen2007}. The lack of the needed photoabsorption data for detecting these near-neutral charge states of gas-phase atomic Fe hinders our ability to determine the Fe abundance, which is critical for advancing our understanding of the ISM.

One important and poorly constrained component of ISM spectra are the atomic Fe L-shell absorption features. Fe L-shell absorption presents a powerful diagnostic of the cold and warm ISM since the  L-shell features fall in the photon-rich part of X-ray spectra and are easy to detect using high-resolution spectroscopy with either gratings or calorimeters. Understanding the Fe L-shell absorption spectrum will also become increasingly important for future high-resolution calorimeter missions, such as the X-ray Astrophysics Recovery Mission (XARM) and Athena.

In response to these data needs, we have initiated a research program that aims at providing L-shell photoabsorption and photoionization data for low-charged iron ions. Here we report on measurements of absolute cross sections for photoabsorption and for single and $m$-fold ($m=1, \ldots, 6$) photoionization of Fe$^+$ ions in the photon energy range 680--920~eV which comprises the iron $2p$ and $2s$ thresholds. In addition, we have carried out theoretical calculations of the Fe$^+$ absorption cross section in the vicinity of the thresholds for the ionization of the iron $L$ shell. We also model the major decay cascades that occur subsequently to the creation of a $2p$ or a $3s$ hole and that determine the distribution of the product charge states.

\section{Previous related work}

The current knowledge from Fe-$2p$ spectroscopy of isolated atoms, low-charged ions, and chemical compounds has been summarized by \citet{Miedema2013}. Previous experimental work on L-shell ionization of iron atoms and ions is scarce. $2p$ absorption in neutral iron atoms has been investigated by \citet{Richter2004} using synchrotron radiation. Similar work has been carried out with other $3d$ elements \citep[see][for a comprehensive review]{Martins2006a}. \citet{Hirsch2012} studied $2p$ photoionization of Fe$^+$ ions in a radio-frequency (RF)  ion trap. Another type of ion trap, i.e., an electron-beam ion trap (EBIT) was used  by \citet{Simon2010a} for $2p$ photoionization of Fe$^{14+}$ ions. Further photoionization experiments with iron ions addressed exclusively the valence shells of Fe$^+$ \citep{Kjeldsen2002c}, of Fe$^{4+}$ \citep{Bizau2006a}, of Fe$^{2+}$ to Fe$^{6+}$ \citep{ElHassan2009}, and of Fe$^{3+}$, Fe$^{5+}$, Fe$^{7+}$ \citep{Gharaibeh2011}.

More distantly related experimental work comprises the spectroscopy of the L-shell emission --- as is observed, e.g., in X-ray spectra of active galactic nuclei \citep{Fabian2009} --- from electron-impact excitation of highly ionized iron ions at an EBIT \citep{Gu2001,Chen2006a,Beiersdorfer2013},  photoexcitation of highly charged iron ions in an EBIT \citep{Epp2007,Bernitt2012,Rudolph2013,Steinbruegge2015}, $3p$ detachment from Fe$^-$ \citep{Dumitriu2010}, and electron-impact ionization and electron-ion recombination  of multiply-charged iron ions in a heavy-ion storage ring \citep[e.g.,][]{Savin2002c,Schippers2010,Bernhardt2014,Hahn2015}. The latter process can be viewed as the time-inverse of photoionization and is therefore very suitable to test corresponding theories \citep{Schippers2002b,Mueller2009a,Mueller2014a}.

\begin{figure*}
\includegraphics[width=\textwidth]{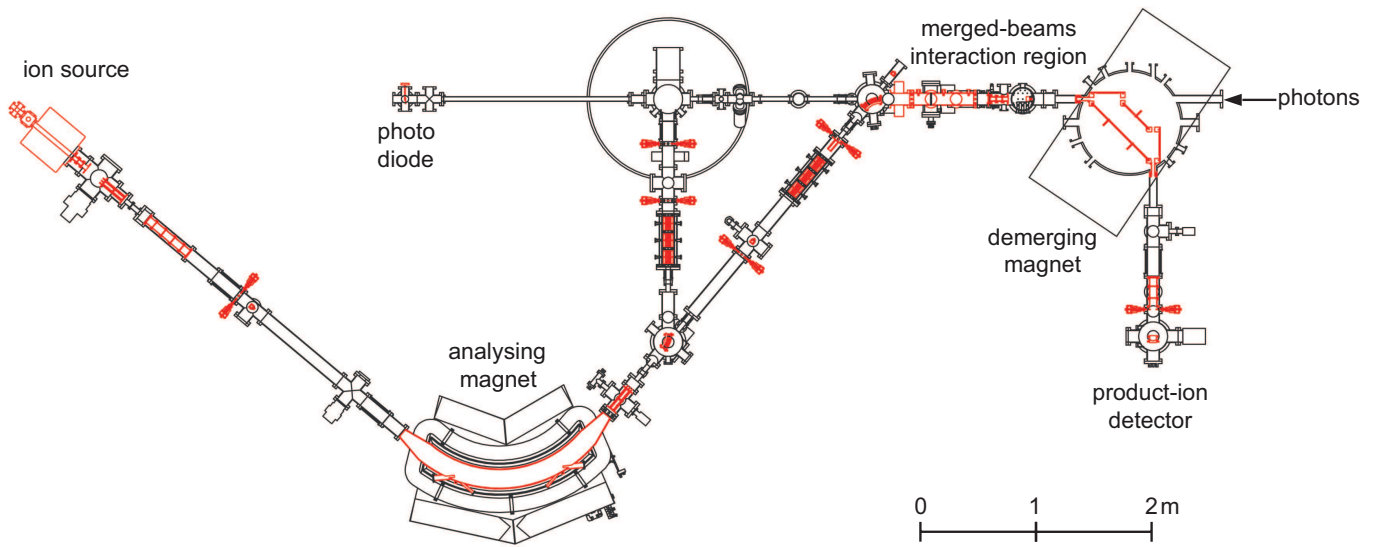}
\caption{\label{fig:exp}\changed{Floor plan of the Photon-Ion Spectrometer setup at PETRA III (PIPE). For a detailed description of all components depicted in this figure see \citet{Mueller2017}.}}
\end{figure*}

Theoretical calculations of cross sections for photoionization of neutral iron and of iron ions have been carried out within the opacity and iron projects \citep{Seaton1994a,Badnell2003b,Badnell2005a,Hummer1993}. Other calculations for Fe$^+$ ions have considered only lower photon energies where excitations of valence shell electrons are important \citep{Nahar1994,Nahar1997a,Hansen2007a,Fivet2012}. Cross sections at higher energies have been provided only for direct ionization of inner-shell electrons \citep{Verner1993a,Verner1995} and, thus, do not account for contributions by resonant ionization processes which are often dominant, particularly in the vicinity of subshell ionization thresholds. A notable exception are the calculations that were performed in support of the $2p$ photoionization measurements with neutral iron atoms by \citet{Richter2004} and which have been described in detail by \citet{Martins2006a}.

\section{Experiment}\label{sec:exp}

The present experiment on photoionization of Fe$^+$ ions was carried out at the {\bf P}hoton-{\bf I}on Spectrometer setup at {\bf PE}TRA III (PIPE, \changed{Fig.~\ref{fig:exp}}). PETRA III is presently the brightest third generation synchrotron light source worldwide. It is operated by DESY in Hamburg, Germany. PIPE is the endstation of the Variable Polarization XUV (extreme UV) Beamline P04 of PETRA III \citep{Viefhaus2013}. The experimental arrangement and procedures have been described previously \citep{Schippers2014,Mueller2017}. The present description is focused on the aspects specific to the Fe$^+$ photoionization experiments described in this publication.

The Fe$^+$ ions for this experiment were produced according to the prescription of \citet{Koivisto1994}, i.e., by evaporating ferrocene (C$_{10}$H$_{10}$Fe) powder and leaking the vapor into an electron-cyclotron-resonance (ECR) ion source.  For the present experiments, a compact 10-GHz permanent-magnet ECR source was used \citep{Trassl1997a}. At room temperature, ferrocene has a sufficiently high vapour pressure of 1~Pa \citep{Monte2006}, such that ferrocene vapour could be easily leaked into the ion source. The ion source was on a potential of +6~kV with respect to ground potential. Positively charged ions were thus accelerated from the ion source into the grounded beam pipe. The beam contained ions of different elements and chemical compounds with different charge states. A dipole magnet was used to separate the ions according to their mass-to-charge ratio. \changed{This magnet is denoted as \lq\lq{}analysing magnet\rq\rq\ in Fig.~\ref{fig:exp}}. For the present experiments a beam consisting of $^{56}$Fe$^+$ ions was selected and transported by ion optical focussing and bending elements to the photon-ion merged-beams interaction region where it was collinearly merged with the photon beam from the monochromator of the XUV beamline. \changed{Product-ions were separated from the primary beam by a second dipole magnet (\lq\lq{}demerging magnet\rq\rq\ in Fig~\ref{fig:exp}). The magnetic field was set such that product ions with a selected charge state were directed onto a single-particle detector where they were counted with nearly 100\% efficiency. The primary ions were collected in a Faraday cup inside the magnet chamber.} Collimating 4-jaw slits are located at the entrance and at the exit of the merged-beams interaction-region. The $^{56}$Fe$^{+}$ ion beam current in the photon-ion interaction region was up to 103~nA with the collimating slits wide open and 26~nA when the collimating entrance slits were narrowed onto the photon beam which had a diameter of 1.2 (2.0)~mm in horizontal (vertical) direction.

\changed{The photon flux was measured with a calibrated photo-diode. It amounted to $2.7-3.9 \times10^{13}$~s$^{-1}$ in the $680-920$~eV photon-energy range at a photon energy spread of about 1~eV.} The calibration of the photon energy scale has been taken over from recent $K$-shell ionization measurements with Ne$^+$ ions \citep{Mueller2017}. This has been possible since both the Ne$^+$ calibration and the present Fe$^+$ measurements were carried out immediately after one another with the same settings of the X-ray photon beam\-line. The only difference in the calibration procedure concerns the correction of the Doppler shift which is determined by the velocity of the ions. This correction is smaller for the slower Fe$^+$ ions than for the faster Ne$^+$ ions. The uncertainty of the experimental photon-energy scale has been discussed comprehensively by \citet{Mueller2017}. For $K$-shell ionization of Ne$^+$ ions, this uncertainty amounted to $\pm200$~meV. The same value is adopted here although the photon energies for iron $L$-shell ionization are somewhat smaller than for $K$-shell ionization of Ne$^+$. This rather large uncertainty is primarily due to the lack of accurate calibration standards in the XUV photon energy range as discussed in detail by \citet{Mueller2017}.

\changed{Absolute cross sections for $m$-fold ionization were derived by normalizing the measured Fe$^{(m+1)+}$ product-ion count-rates on the primary ion current, on the photon flux, and on the geometrical beam overlap. The geometrical overlap was determined from beam profiles that were measured with slit scanners at three different locations along the photon-ion merged-beams interaction region. The length of the interaction-region (49.5~cm) was fixed by voltage labelling as detailed by \citet{Schippers2014}.} These authors also discussed the error budget of cross section measurements at PIPE. The systematic uncertainty of the absolute cross section scale was found to be 15\% at a 90\% confidence level. Statistical uncertainties are typically small and reach at most a few percent. Only at very low signal rates can the statistical uncertainties be higher. \changed{In principle, it is also possible that there are contributions by sequential two-photon ionization to the measured one-photon ionization cross sections. However, a detailed estimation (see appendix) reveals, that contribution by sequential two-photon ionization are negligible in all measured ionization channels.}

The ground level of Fe$^+$ is the $3d^6(^5D)\,4s\;a\,^6D_{9/2}$ level\footnote{Here and throughout, closed inner-subshells are not always included in the notation.}. There are 62 excited levels belonging to the $3d^7$, $3d^6\,4s$, and $3d^5\,4s^2$ configurations with the same (even) parity that are also energetically below the first excited level of the opposite (odd) parity, i.e., the $3d^6(^5D)\,4p\;z\,^6D^\circ_{9/2}$ level at an excitation energy of 4.768~eV \citep{Kramida2015}. \citet{Bautista2015} performed large-scale atomic structure calculations of the 52 energetically lowest levels of Fe$^{+}$. The resulting lifetimes of these levels vary from 0.24~ms to $\sim$15,000~s. This is much longer than the $\sim 70$-$\mu$s flight time of the ions from the ion source to the photon-ion interaction region. Consequently, it must be assumed that the ion beam contained a mixture of Fe$^+$ ions in the ground level and in long-lived metastable levels. Generally, the fractional populations of the various levels depend on the plasma conditions in the ion source. \citet{Kjeldsen2002c} used a hot-filament plasma ion source and inferred a metastable fraction of 10\% in their Fe$^+$ ion beam by assuming a Boltzmann distribution with $k_BT = 0.12$~eV over the Fe$^+$ energy levels. However, in an ECR ion source, the plasma temperature is generally higher and a statistical distribution over the levels of a given term is often an appropriate assumption \citep[see, e.g.,][]{Mueller2015c} which we adopt also here.

\section{Theory}\label{sec:theo}

In order to analyze the experimental Fe$^+$ photoionization spectra we have performed two different variants of atomic-structure calculations. First, Hartree-Fock calculations including relativistic extensions (HFR) have been performed using the \textsc{Cowan} code \citep{Cowan1981} and the extended Fano theory \citep{Fano1961, Mies1968a, Martins2001} resulting in cross sections for \emph{direct} and \textit{resonant} photoabsorption of Fe$^+$. Configuration interaction (CI) has been included in the ground levels and the photoexcited levels as described in more detail below.  Second, fully relativistic Multi-Configuration Dirac-Fock (MCDF) calculations have been carried out using the \textsc{Grasp} \citep{Joensson2007} and \textsc{Ratip} \citep{Fritzsche2012a} codes resulting in cross sections specifically for \textit{resonant} photoabsorption. In these calculations, deexcitation cascades (Tab.~\ref{tab:2p3dexci}) have been tracked by taking the dominant radiative and nonradiative processes into account and arriving at distributions of product charge states as in previous similar work on oxygen anions \citep{Schippers2016a} and neutral neon \citep{Stock2017}. It should be noted that extensive cascade calculations have also been performed by \citet{Kucas2015} for inner-shell photoionization for a number of astrophysically relevant ions. However, these authors did not consider iron ions.

\subsection{HFR calculations}\label{sec:hfr}

\begin{deluxetable*}{ll}
\tablecaption{\label{tab:2p3dexci}\changed{Major paths of sequential Auger decays to the various final product-charge states $q$ after resonant $2p \to 3d$ excitation of Fe$^+$ as considered in the present MCDF calculations.  Unfilled inner subshells are in bold face. The charge states $q=6$ and $q=7$ cannot be reached by stepwise emission of only one electron per step. The production of these charge states requires multi-electron processes to occur such as double or triple Auger processes \citep[see, e.g.,][]{Mueller2015a} or double-shake processes \citep[see, e.g.,][]{Schippers2016a}.}}
\tablehead{
\colhead{$q$} &
\colhead{Decay path}
}
\startdata
2 & $\boldsymbol{2p^5} 3s^2 3p^6 3d^7 4s \to  2p^6 3s^2 3p^6 3d^5 4s$ \\
 & \\
3 & $\boldsymbol{2p^5} 3s^2 3p^6 3d^7 4s \to 2p^6 3s^2 \boldsymbol{3p^5} 3d^6 4s \to 2p^6 3s^2 3p^6  3d^4 4s$ \\
3 & $\boldsymbol{2p^5} 3s^2 3p^6 3d^7 4s \to 2p^6 3s^2 \boldsymbol{3p^5} 3d^6 4s \to 2p^6 3s^2 3p^6  3d^5$ \\
3 & $\boldsymbol{2p^5} 3s^2 3p^6 3d^7 4s \to 2p^6 3s^2 \boldsymbol{3p^5} 3d^7 \mathit{\phantom{4s}}\to 2p^6 3s^2 3p^6 3d^5$ \\
 & \\
4 & $\boldsymbol{2p^5} 3s^2 3p^6 3d^7 4s \to 2p^6 3s^2 \boldsymbol{3p^4} 3d^7 4s \to 2p^6 3s^2 \boldsymbol{3p^5} 3d^5 4s \to 2p^6 3s^2 3p^6 3d^3 4s$ \\
4 & $\boldsymbol{2p^5} 3s^2 3p^6 3d^7 4s \to 2p^6 3s^2 \boldsymbol{3p^4} 3d^7 4s \to 2p^6 3s^2 \boldsymbol{3p^5} 3d^5 4s  \to 2p^6 3s^2 3p^6 3d^4$ \\
4 & $\boldsymbol{2p^5} 3s^2 3p^6 3d^7 4s \to 2p^6 3s^2 \boldsymbol{3p^4} 3d^7 4s \to 2p^6 3s^2 \boldsymbol{3p^5} 3d^6\mathit{\phantom{4s}}  \to 2p^6 3s^2 3p^6 3d^4$  \\
4 & $\boldsymbol{2p^5} 3s^2 3p^6 3d^7 4s \to 2p^6 \boldsymbol{3s}^\mathit{\phantom{2}} \boldsymbol{3p^5} 3d^7 4s \to 2p^6 \boldsymbol{3s}^\mathit{\phantom{2}} 3p^6 3d^5 4s \to 2p^6 3s^2 3p^6 3d^4$ \\
 & \\
5 & $\boldsymbol{2p^5} 3s^2 3p^6 3d^7 4s \to 2p^6 \boldsymbol{3s}^\mathit{\phantom{2}} \boldsymbol{3p^5} 3d^7 4s \to 2p^6 \boldsymbol{3s}^\mathit{\phantom{2}} 3p^6 3d^5 4s \to 2p^6 3s^2 \boldsymbol{3p^5} 3d^5 \mathit{\phantom{4s}} \to 2p^6 3s^2 3p^6 3d^3$ \\
5 & $\boldsymbol{2p^5} 3s^2 3p^6 3d^7 4s \to 2p^6 \boldsymbol{3s}^\mathit{\phantom{2}} \boldsymbol{3p^5} 3d^7 4s \to 2p^6 3s^2 \boldsymbol{3p^4} 3d^6 4s \to 2p^6 3s^2 \boldsymbol{3p^5} 3d^4 4s \to 2p^6 3s^2 3p^6 3d^3$ \\
5 & $\boldsymbol{2p^5} 3s^2 3p^6 3d^7 4s \to 2p^6 \boldsymbol{3s}^\mathit{\phantom{2}} \boldsymbol{3p^5} 3d^7 4s \to 2p^6 3s^2 \boldsymbol{3p^4} 3d^6 4s \to 2p^6 3s^2 \boldsymbol{3p^5} 3d^4 4s \to 2p^6 3s^2 3p^6 3d^2 4s$ \\
5 & $\boldsymbol{2p^5} 3s^2 3p^6 3d^7 4s \to 2p^6 \boldsymbol{3s}^\mathit{\phantom{2}} \boldsymbol{3p^5} 3d^7 4s \to 2p^6 3s^2 \boldsymbol{3p^4} 3d^6 4s \to 2p^6 3s^2 \boldsymbol{3p^5} 3d^5 \mathit{\phantom{4s}} \to 2p^6 3s^2 3p^6 3d^3$ \\
5 & $\boldsymbol{2p^5} 3s^2 3p^6 3d^7 4s \to 2p^6 \boldsymbol{3s}^\mathit{\phantom{2}} \boldsymbol{3p^5} 3d^7 4s \to 2p^6 3s^2 \boldsymbol{3p^4} 3d^7 \mathit{\phantom{4s}}\to 2p^6 3s^2 \boldsymbol{3p^5} 3d^5 \mathit{\phantom{4s}}\to 2p^6 3s^2 3p^6 3d^3$ \\
\enddata
\end{deluxetable*}

For the calculation of the ground-level energies, CI of the configurations\footnote{The \lq{}+\rq\ sign denotes configuration interaction (CI).} $3d^5 4s^2 + 3d^6 4s + 3d^{7}$ has been taken into account. Cross sections for photoabsorption have been calculated for $2p$ and $2s$ core excitations from the initial levels into $2p^5 3s^2 (3d^6 4s^2 + 3d^7 4s + 3d^8)$ and  $2s^1 2p^6 3s^2(3d^5 4s^2 4p + 3d^6 4s 4p + 3d^7 4p)$ levels, respectively. In addition, the corresponding configurations for direct $2p$ and $2s$ ionizations have been included in the calculations as well.

Auger transition rates have been evaluated by including the final ionic Fe$^{2+}$  configurations. For the $2p$ core-hole levels, we have taken into account the configurations  $3s^2 3p^4 3d^{8-k} 4s^k \epsilon(s,d)$, $3s^2 3p^6 3d^{6-k} 4s^k \epsilon(s,d)$, and  $3s^2 3p^5 3d^{7-k} 4s^k  \epsilon(p,f)$ with $k=0-2$. Here $\epsilon$ represents the outgoing electron. Decay channels involving double $3s$ holes have not been considered, since the corresponding Auger transitions are expected to be of minor importance.

For the $2s$ core-hole levels, the number of final ionic configurations had to be restricted to keep the calculations manageable. Nevertheless, all participator\footnote{\changed{In a participator decay the photoexcited electron will take part in the subsequent Auger decay.}} channels have been accounted for. A calculation including only the $4p$ participator configurations underestimated the Auger transition rates of the $2s\,2p^63s^23p^6 3d^{(6-k)} 4s^k 4p$ resonances by 1--2 orders of magnitude. The most important final channels are those belonging to the $4p$ spectator\footnote{\changed{In a spectator decay the photoexcited electron does not take part in the subsequent Auger decay.}}  configurations. Hence, the spectator channels $2p^5 3s^2 3p^6 3d^{6-k} 4s^{k} 4p\, \epsilon(pf)$ have been included in addition.  The number of resonances calculated amounts to 300--400 for every angular momentum $J$ and the number of continuum states is up to 17,000 final levels per $J$.

The typical line widths calculated from the Auger transition rates are in the range 300--400 meV for the $2p$ core-hole levels. They are significantly larger (up to some eV) for the $2s$ core-hole levels because these can decay via fast $L_1L_{2,3}M$ Coster-Kronig processes\footnote{An $XYZ$ Auger process may occur subsequent to the creation of a hole in the $X$ shell which is filled by an electron from the $Y$ shell, autoionizing an electron from the $Z$ shell.  This is called a Coster-Kronig process when the vacancy is filled by an electron in a higher subshell of the same principal shell.}.

\changed{The HFR calculations result in cross sections for non-resonant and resonant photoabsorption. Deexcitation cascades of the core excited states that lead to the formation of the various product charge-states are only considered in the MCDF calculations as explained next.}

\subsection{MCDF calculations}\label{sec:mcdf}

The absorption cross section due to resonant excitation of a $2p$ or $2s$ core electron was computed based on wave functions  for the $3d^6 4s$ ground configuration and the $2p^5 3d^7 4s$ excited configuration in single-configuration approximation giving rise to a total of 63 and 213 fine-structure levels, respectively. The contribution of the $2p^5 3d^6 4s^2$ configuration that corresponds to resonant $2p \rightarrow 4s$ excitations was found to be very small and was hence neglected in all subsequent computations.

\begin{figure}[t]
\includegraphics[width=0.9\figurewidth]{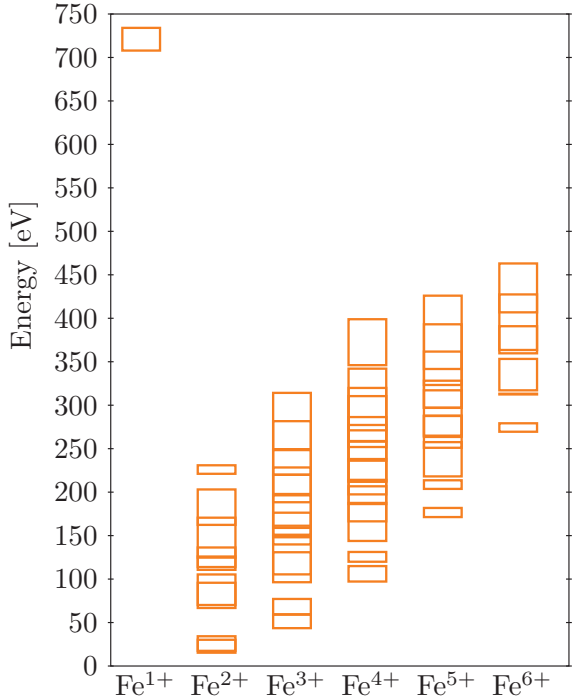}
\caption{\label{fig:levelDiagram}
Computed level diagram of all configurations that are relevant in the Auger decay of the $2p^5\,3s^2\,3p^6\,3d^7\,4s$ configuration in Fe$^{1+}$.  All configurations that arise due to single Auger processes are shown, where the energy range covered by the fine-structure levels of each configuration is represented by a rectangle. The top left rectangle represents the levels that belong to the $2p^5\,3s^2\,3p^6\,3d^7\,4s$ configuration of Fe$^+$. Single Auger decay corresponds to a rightward transition to the next higher charge state into a level at lower energy. It should be noted that the energetically lowest level of Fe$^{6+}$ is above the highest Fe$^{2+}$ level. This implies, that Fe$^{6+}$ cannot be produced by a cascade consisting only of single Auger transitions.}
\end{figure}

\begin{deluxetable*}{lllllllll}
\tablecaption{\label{tab:Xsec}Measured cross sections $\sigma_m$ (Fig.~\ref{fig:all}) for $m$-fold photoionization of Fe$^+$ ions, resulting sum cross section $\sigma_\Sigma$ (Eq.~\ref{eq:sigmasum}, Fig.~\ref{fig:absorp}), and mean product charge-state $\overline{q}$ (Eq.~\ref{eq:qmean}, Fig.~\ref{fig:fractions}b). The numbers in parentheses provide the statistical experimental uncertainties. The systematic uncertainties of the energy and cross-section scales are $\pm0.2$~eV and $\pm15\%$ (at 90\% confidence level), respectively.}
\tablehead{
    \colhead{Energy~(eV)} &
    \colhead{$\sigma_{1}$~(Mb)\tablenotemark{a}}&
    \colhead{$\sigma_{2}$~(Mb)}&
    \colhead{$\sigma_{3}$~(Mb)}&
    \colhead{$\sigma_{4}$~(Mb)}&
    \colhead{$\sigma_{5}$~(Mb)}&
    \colhead{$\sigma_{6}$~(Mb)}&
    \colhead{$\sigma_{\Sigma}$~(Mb)} &
    \colhead{$\overline{q}$}
    }
\startdata
680.911 & 0.0299(45) & 0.1431(24) & 0.0476(20) & 0.0042(6)  & 0.0003(2)  & 0.00004(6)  & \phn0.2251(55) & 3.120(24)\\
700.137 & 0.0162(42) & 0.1313(22) & 0.0544(21) & 0.0065(7)  & 0.0008(3)  & 0.00006(6)  & \phn0.2091(53) & 3.336(27)\\
708.348 & 2.555(13)	 & 3.259(11)  &	4.617(19)  & 0.5094(65) & 0.0402(20) & 0.0010(12) &     10.982(26)  & 3.292(2)\\
715.758	& 0.0806(47) & 0.1956(27) &	0.1373(33) & 0.0180(12) & 0.0010(3)  & 0.00005(2)  & \phn0.4323(65) & 3.221(16)\\
720.364 & 0.4967(71) & 0.8635(56) & 1.1587(94) & 0.1434(34) & 0.0103(10) & 0.00039(11) & \phn2.673(14)  & 3.367(5)\\
738.188 & 0.0411(43) & 0.4175(38) & 0.9887(87) & 0.2957(48) & 0.0216(14) & 0.00066(15) & \phn1.765(12)  & 3.909(6)\\
800.072 & 0.0278(40) & 0.2750(30) & 0.6700(70) & 0.3954(55)	& 0.0291(16) & 0.00120(19) & \phn1.399(10)   & 4.088(8)\\
855.946 & 0.0240(40) & 0.2400(29) &	0.6524(69) & 0.4662(60) & 0.0603(23) & 0.00265(27) & \phn1.446(11)   & 4.207(8)\\
900.005 & 0.0181(41) & 0.2095(27) &	0.5657(64) & 0.4171(58) & 0.0867(29) & 0.00390(32) & \phn1.302(10)   & 4.266(9)\\
\enddata
\tablenotetext{a}{1~Mb = $10^{-18}$~cm$^2$}
\tablenotetext{}{(This table is available in its entirety in machine-readable form.)}
\end{deluxetable*}

After the creation of an inner-shell hole, the deexcitation of the core excited levels generally proceeds by Auger and radiative transitions. In principle, higher order processes like multiple Auger processes where more than one electron is emitted at a time \citep{Mueller2015a} can play a role. Here we only consider de-excitation pathways by sequential Auger decay \changed{(for examples, see Tab.~\ref{tab:2p3dexci})}. These include all electronic configurations that arise due to single Auger decay processes emerging from the core-hole excited $2p^5 3d^7 4s$ configuration. The excitation energies of all configurations that emerge in this way are shown in Fig.~\ref{fig:levelDiagram}. However, it must be noted, that the population of many levels of these configurations by sequential single Auger processes is energetically not possible. Therefore, when double Auger processes as well as shake-up transitions are neglected, only ions up to Fe$^{5+}$ can be produced since the populated levels with the highest energy in the cascade pathways belong to the $3s^{-2}$ configuration in Fe$^{2+}$. Conversely, the production of Fe$^{6+}$ ions would require at least one step involving a double Auger or shake-up process. The total non-radiative decay width of the 213 fine-structure levels of the $2p^5 3d^7 4s$ configuration varies from $350\, \mathrm{meV}$ to about $500\, \mathrm{meV}$, which is expected to be somewhat overestimated due to the non-orthogonality of the underlying orbital basis sets for the initial and final wave function expansions.

We model in a similar manner the Auger cascades that emerge from  Fe$^{2+}$ created by single-photon single-electron ionization processes. However, in this case it becomes necessary to account not only for the dominant ionization of $2p$ electrons but also competing ionization processes from higher (sub)-valence shells. There\-fore, the Auger cascades that emerge from $2p^{-1} 3d^6 4s$, $3s^{-1} 3d^6 4s$, and $3p^{-1} 3d^6 4s$ holes have been  modeled independently. Based on the computed cross sections for direct photoionization, these results are then combined to obtain the total ion yields at a specific photon energy. As an aside, we note that the observed production of any stable Fe$^{2+}$ ions is either through radiative stabilization of the initial autoionizing state or due to the valence ionization of either $3d$ or $4s$ electrons.

\begin{figure*}
\centering{\includegraphics[width=0.8\textwidth]{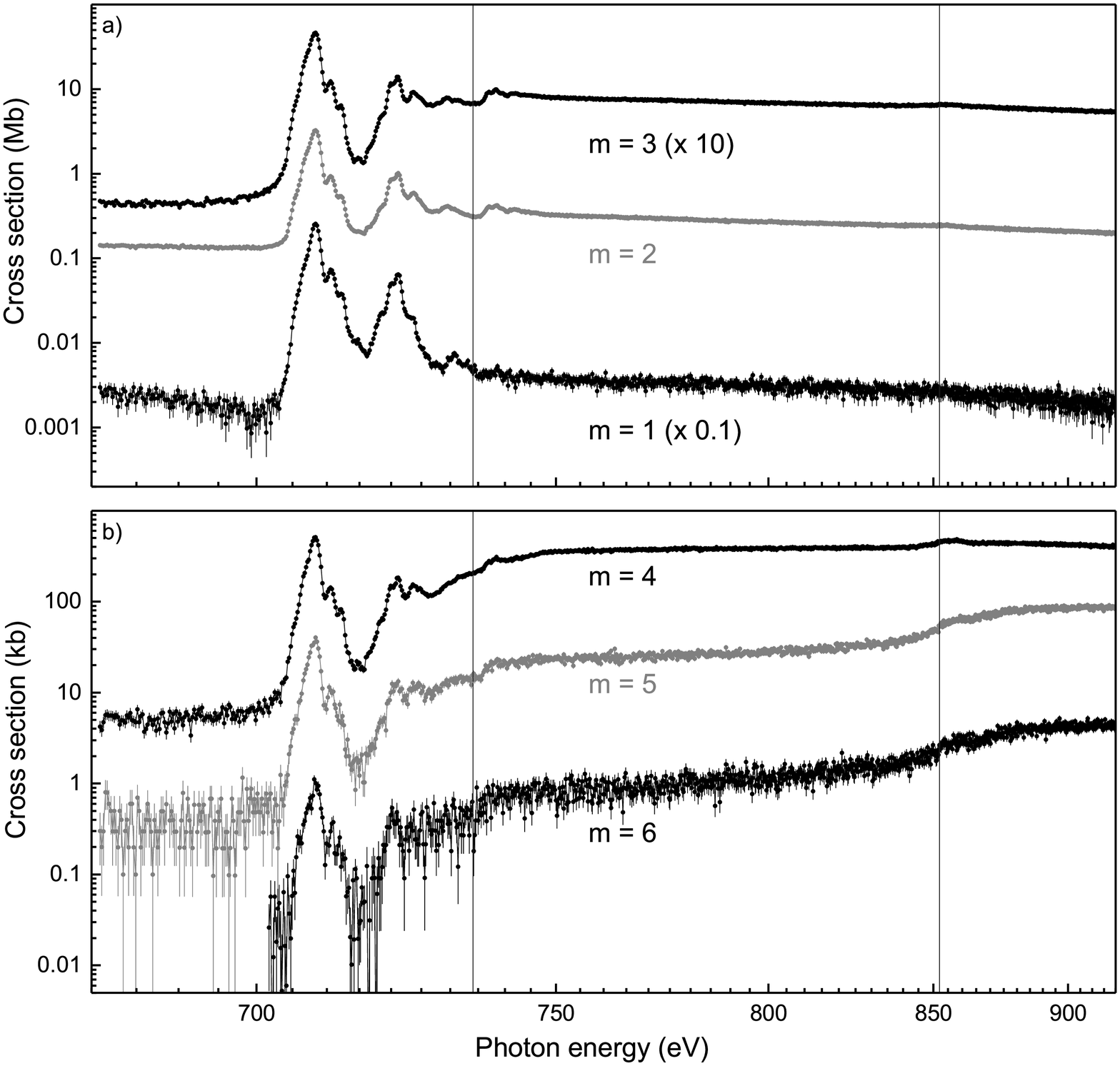}}
\caption{\label{fig:all}Measured cross sections $\sigma_m$ for $m$-fold photoionization of Fe$^+$. The experimental photon energy spread was about 1~eV. The cross sections $\sigma_1$ and $\sigma_3$ were multiplied by factors of 0.1 and 10, respectively, for a clearer presentation
of the data. Different cross-section units are used in the panels a) and b).  The vertical lines mark the $2p$ and $2s$ subshell ionization thresholds at 734.1 and 852.1~eV as calculated by \citet{Verner1993a}. The photon-energy ($E_\mathrm{ph}$) axis is compressed
towards higher energies \changed{using $x=\log(E_\mathrm{ph}/\mathrm{eV}-600)$ for the abscissa}
in order to provide a better view of the resonance structures at lower energies. }
\end{figure*}

This cascade model gives rise to several thousand fine-structure levels for the intermediate charge states, and hence millions of Auger transitions between these levels. In order to keep the calculations of the Auger transition rates feasible, it was first necessary to constrain the size of the Auger matrices. Therefore, all wave functions were computed in single-configuration approximation. This approach to the generation of the wave functions has been detailed by \citet{Buth2017}. It neglects effects due to configuration interaction that become crucial for the description of shake-processes such as in the work of \citet{Andersson2015} and \citet{Schippers2016a}. In the next step of this approach, the transition rates between fine-structure levels of the configurations were averaged assuming a statistical population to obtain an average transition rate between configurations as it is detailed by \citet{Buth2017}. However, this approach yields results that are very similar to the previous computations by \citet{Kaastra1993} that do not reproduce the experimental findings very well. Therefore, we adopted another approach that tracked the full decay tree between fine-structure levels based on the single Auger transition rates computed in single-configuration approximation. In this approach, the highly non-statistical population of the fine-structure levels of one configuration due to the resonant $2p \rightarrow 3d$ excitation or direct $2p$ ionization is fully accounted for. Nevertheless, it still neglects radiative losses since the associated radiative transitions were assumed to be much slower than the Auger processes. However, the comparison with the experimental findings below suggests that this assumption is well justified.

\section{Results and discussion}\label{sec:res}

Our measured absolute cross sections $\sigma_m$ for $m$-fold photoionization ($m=1,\ldots,6$) of singly charged iron ions are displayed in  Fig.~\ref{fig:all} as a function of photon energy. The measured cross section values span almost four orders of magnitude ranging from less than 0.1~kb (1~b = 1~barn = $10^{-24}$~cm$^2$) to about 5~Mb (see also Tab.~\ref{tab:Xsec}).  The fact that even the weak cross section for the production of Fe$^{7+}$ could be measured with reasonable statistical uncertainties demonstrates the extraordinary experimental sensitivity of the PIPE setup.

All measured cross sections exhibit the resonance structure below the $2p$ ionization threshold. These resonances are associated with resonant photoexcitation of a $2p$ electron into the $3d$ and higher subshells  and converge towards the $2p$ ionization threshold of ground-level Fe$^+$ which is expected to occur at a photon energy of 734.1 eV \citep{Verner1993a}. This threshold cannot clearly be discerned in the experimental cross sections because the photoionization resonances blend together. The $2s$ threshold at 852.1 eV, according to the calculations of \citet{Verner1993a}, is most pronounced in the cross sections for four-, five-, and sixfold ionization.

Clearly, the shapes of the partial cross sections as a function of photon energy differ from one another. This contradicts the assumption of \citet{Hirsch2012} that each partial cross section is proportional to the absorption cross section. Above the $2p$-ionization threshold, the cross sections for single, double, and triple ionization decrease monotonically with increasing photon energy. Conversely, the cross sections for the production of higher charge states increase, in particular, beyond the $2s$-ionization threshold. As expected, $2s$ ionization leads, on average, to the production of more highly charged ions than an initial $2p$ ionization.

Further differences between the partial cross sections concern the resonance strengths and line shapes. Resonance strengths vary because the branching ratios for the decay of the associated multiply excited states into the various final charge states are specific for each resonance level. The line shape of the lowest-energy resonance in the single-ionization cross-section \changed{($m=1$ in Fig.~\ref{fig:all})} exhibits an asymmetry which manifests itself as a dip of the cross section immediately below the onset of the resonance. Such a dip does not occur in the other partial cross sections. The asymmetric line shape in $\sigma_1$ is an indication of quantum mechanical interference between resonant ionization and direct ionization \citep{Fano1965} which is a common phenomenon in the photoionization of atoms \citep{Madden1963a,Martins2006a} and ions \citep{Kjeldsen2006a,Schippers2016}.

Here the interference is between ionization via the formation of an intermediate $2p^5\,3s^2\,3p^6\,3d^7\,4s$ resonance with subsequent LMM-Auger decay and direct $3d$ ionization into a $2p^6\,3s^2\,3p^6\,3d^5\,4s$ level. It should be noted that the fast LMM-Auger decay involving two $3d$ electrons is the dominant pathway to single ionization. If a $3p$ electron is involved the resulting $3p$ hole will be rapidly filled by a subsequent Super-Coster-Kronig\footnote{Type of Auger process in which the vacancy is filled by an electron from a higher subshell of the same principal shell and the emitted electron  (the \lq{}Auger electron\rq) also belongs to the same principal shell (here the M shell).} decay \citep{McGuire1972,Gharaibeh2011} leading to the emission of a second $3d$ electron, and, thus to the formation of a triply charged ion. The limited number of channels available for single-ionization facilitate the observation of the Fano line shape in the single-ionization cross section. Many more decay channels are available for the formation of the higher charge states. These contribute mostly incoherently to the respective partial cross sections and this conceals the asymmetries of the few decay channels which are subject to strong interference effects.
\changed{Despite the fact that the number of single-ionization channels is limited, an unambiguous extraction of Fano asymmetry parameters from the measured single-ionization cross-section by peak fitting is prohibitive because the number of fine-structure levels of the $2p^6\,3s^2\,3p^6\,3d^6\,4s$ ground configuration and the $2p^5\,3s^2\,3p^6\,3d^7\,4s$ intermediate configuration (amounting to 63 and 213, respectively) is too large for a meaningful individual treatment of all associated resonances.}

\subsection{Absorption cross section}

Since we have measured all the significant ionization channels, the sum of all measured partial cross sections, i.e.,
\begin{equation}
\label{eq:sigmasum}
\sigma_\Sigma = \sum_{m=1}^6\sigma_m,
\end{equation}
represents the Fe$^+$ photoabsorption cross-section provided that photon-scattering can be neglected. This assumption is justified since the fluorescence yields from the multiply excited resonance states are generally negligible for a comparatively light element such as iron \citep{McGuire1972}.

\begin{figure}
\includegraphics[width=\figurewidth]{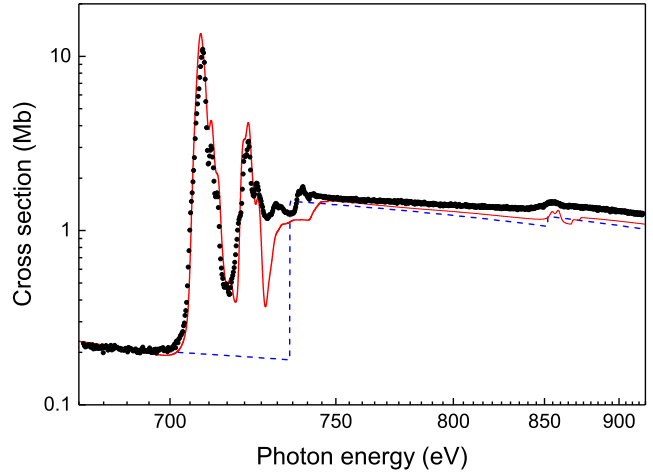}
\caption{\label{fig:absorp}Experimental cross section sum $\sigma_\Sigma$ (Eq.~\ref{eq:sigmasum}, filled circles), present theoretical HFR cross section for photoabsorption (full red line, statistical average of all $3d^6(^5D)4s\;^6D_J$ levels, convoluted with a Gaussian of 1.0 eV FWHM), and the theoretical cross section for photoabsorption from \citet[][short-dashed blue line]{Verner1993a}.  As in Fig.~\ref{fig:all}, the energy axis is compressed towards higher energies in order to provide a better view of the resonance structures at lower energies.}
\end{figure}

Figure~\ref{fig:absorp} displays the experimental \lq{}absorption cross-section\rq\ $\sigma_\Sigma$  as a function of photon energy. The cross section is compared with the present HFR calculations and the theoretical results of \citet{Verner1993a}. The latter comprise only direct photoionization processes. Consequently, the resulting cross sections exhibit edge- or step-like features at the photon energies which coincide with subshell ionization thresholds. In the present photon-energy range these are the $2p$ and $2s$ thresholds for ground-level Fe$^+$ at 734.1~eV and  852.1~eV, respectively. In the energy ranges where there are no resonance features the cross section of Verner et al.\ agrees with the experimental data to within the $\pm$15\% experimental uncertainty at energies up to 817.5~eV. At the energies beyond the $2s$ threshold the deviation is somewhat larger reaching up to 19\% at 918~eV. This discrepancy may be due to the neglect of direct multiple ionization (e.g., simultaneous ionization of a $2p$ and a $4s$ electron) in the theoretical cross section.

\begin{figure}
\includegraphics[width=\figurewidth]{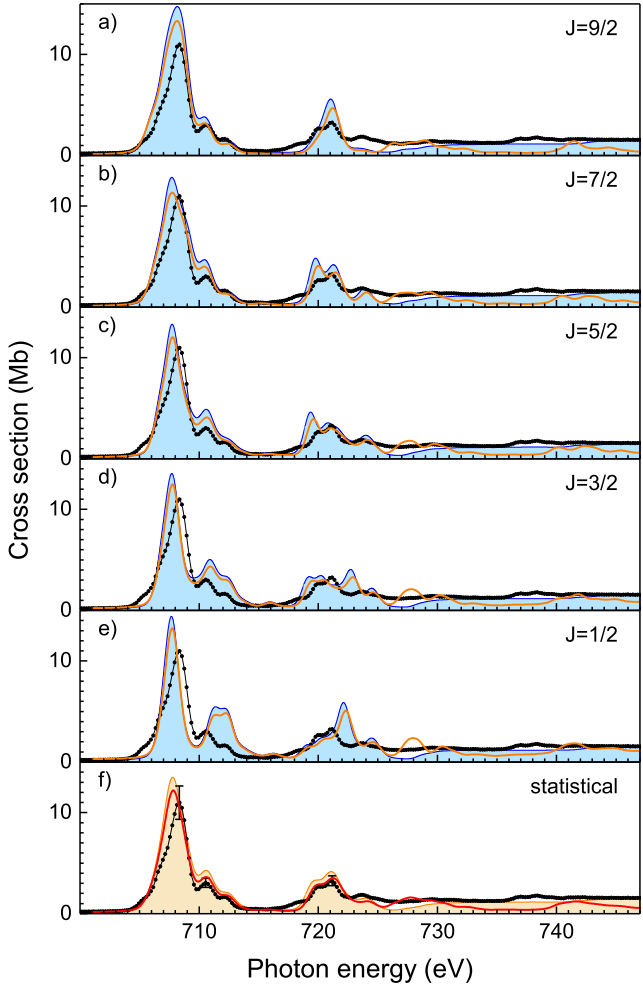}
\caption{\label{fig:theo} a) to e) Comparison of the experimental absorption cross section (circles) with the present HFR (shaded curves) and MCDF (full lines) theoretical results  for different $3d^6(^5D)4s\;^6D_J$ levels of the Fe$^+$ primary beam. f) Statistical average of all $J$-levels belonging to the $3d^6(^5D)4s\;^6D$ ground term. The experimental error bars in panel f) indicate the $\pm15\%$ uncertainty of the absolute cross section scale. All theoretical cross sections were convoluted with a Gaussian of a FWHM of 1.0~eV. The energy axes of the HFR and MCDF calculations were shifted by -2.8 and +1.0~eV, respectively.}
\end{figure}

As far as direct ionization is concerned, the present HFR calculations show better agreement with the experimental data than the cross section of \citet{Verner1993a}. In addition to direct ionization, the present HFR photoabsorption cross-sections include contributions by resonant photoionization processes as discussed in Section~\ref{sec:theo}. The energy scale of the HFR calculations has been shifted slightly, as discussed below, but there is a remarkable agreement in resonance widths and strengths (see also Fig.~\ref{fig:theo}) considering the level of complexity of the atomic-structure of the inner-shell excited Fe$^+$ ion.  The two strongest resonance groups with their respective maxima at about 708.3 and 721.1 eV are associated with $2p\to3d$ excitations leaving a $2p_{3/2}$ or a $2p_{1/2}$ core hole behind, respectively. This 12.8-eV fine-structure splitting is practically the same as for neutral iron \citep{Richter2004}. It is well reproduced by the present calculations.

Panels a) to e) of Fig.~\ref{fig:theo} show our HFR and MCDF cross sections for photoabsorption of Fe$^+$ ions separately for the different $3d^6(^5D)4s\;^6D_J$ levels, i.e., for $J=9/2, 7/2, 5/2, 3/2$, and $1/2$. Panel f) displays the statistical average  of all these cross sections, i.e., the weighted cross-section sum with the individual weights equalling 2$J$+1. For the comparison with the experimental result the HFR cross sections were shifted by 2.8 eV towards lower photon energies and the MCDF cross sections were shifted by 1.0 eV towards higher photon energies. All theoretical cross section data were convoluted with a Gaussian of a full width at half maximum (FWHM) of 1.0 eV in order to account for the experimental energy spread. The statistically averaged cross sections agree best with the experimental findings. The comparisons in Figure~\ref{fig:theo} suggest that the gross resonance structure in the photoabsorption cross section does not depend drastically on the level populations in the primary ion beam, in particular, when comparing the pure ground-level cross section in Fig.~\ref{fig:theo}a with the statistically averaged one in Fig.~\ref{fig:theo}f.

\begin{figure}
\includegraphics[width=\figurewidth]{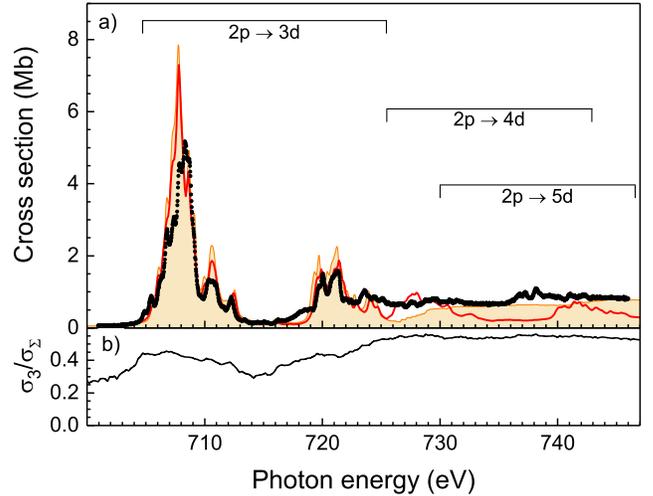}
\caption{\label{fig:hires} a) High-resolution measurement of the cross section for triple ionization of Fe$^+$ ions (filled circles, 100 meV experimental photon-energy spread). Also shown are theoretical absorption cross sections (statistical average over all $3d^6(^5D)4s\;^6D_{J}$ levels convoluted with a Gaussian of a FWHM of 0.1 eV) using the HFR method (shaded curve) and the MCDF method (full red line). The theoretical curves have been  multiplied by the experimental energy-dependent scale factor $\sigma_3/\sigma_\Sigma$. This scale factor is shown in panel b). The energy ranges where resonances associated with $2p\to nd$ excitations with $n=3,4,5$ occur according to our MCDF calculations are indicated.}
\end{figure}

Apart from the different overall energy shifts, both theoretical results largely agree with one another.  At energies below 725~eV the shapes of the resonance structures are almost identical. There the MCDF cross sections agree better in magnitude with the experimental cross section than the HFR cross sections which are larger than the MCDF cross sections by about 10\%. The differences between the theoretical results at energies above 725~eV are due to an underestimation of direct ionization in the MCDF calculations and the neglect of $2p \to 4d$  and higher excitations in the HFR calculations.

In order to better resolve the photoionization resonance structure, a high-resolution scan of the largest partial cross section, i.e., $\sigma_3$ (Fig.~\ref{fig:all} and Tab.~\ref{tab:Xsec}), was performed. Figure~\ref{fig:hires} compares the experimental result with the scaled theoretical absorption cross sections. This scaling was applied, since a rigourous theoretical treatment of the deexcitation cascades leading to the production of Fe$^{4+}$ ions is computationally prohibitive (see below). The scaled HFR and MCDF cross sections reproduce many details of the experimentally observed resonance structure associated with $2p\to3d$ excitations. There is a good agreement between experiment and both theories for the widths of the resonance structures. Agreement with the $2p\to3d$ resonance positions was obtained by shifting the HFR and MCDF energy scales by -2.8 and 1.0 eV, respectively, as already mentioned above. According to our calculations individual resonance widths of the $2p$ excited levels amount to typically 0.4~eV. This is four times larger than the experimental photon energy spread in Fig.~\ref{fig:hires}. Nevertheless, most of the experimentally observed resonance structures are considerably broader, because they are blends of several individual resonances.

The MCDF calculations also account for $2p\to4d$ and $2p\to5d$ resonant excitations (Fig.~\ref{fig:hires}), which were not included in the HFR calculations.  The energy separation between the $4d$ and $5d$ subshells is smaller than the $2p_{1/2}-2p_{3/2}$ fine-structure splitting of the $2p$ core hole. Therefore, the two calculated resonance structures in the energy ranges 726--733~eV and 740--742~eV are both blends of resonances associated with $2p\to4d$ and $2p\to5d$ excitations. The corresponding resonance structures are also visible in the experimental cross section, albeit at energies lower by about 3 eV. The MCDF resonance strengths for these resonances seem to be larger than the experimental ones considering the size of the cross section for direct ionization. These discrepancies are most likely due to the approximations made in the MCDF calculations. The MCDF resonance strengths could probably be reduced by taking more configurations into account in the optimization of the atomic wave functions.

Resonances that are formed by $2s$ excitation produce the broad structure ranging from about 850 to 865~eV in Fig.~\ref{fig:absorp}. The calculated widths of these resonances amount to typically 3~eV. Since the $2s$ hole is rapidly filled by a fast $L$-shell Coster-Kronig transition, this is much larger than the $\sim$0.4~eV widths of the resonances associated with $2p$ excitation. Similar to our HFR treatment of the $2p$ excitation, our HFR calculations for the $2s$ excitation do not take into account excitations to higher subshells than the $4p$ subshell. This produces the artificial dip in the calculated HFR cross section near 865~eV. These calculations predict  that the $2s$ ionization threshold occurs at 867~eV. This is a significantly higher value than the threshold energy calculated by \citet{Verner1993a}.

\begin{figure}
\includegraphics[width=\figurewidth]{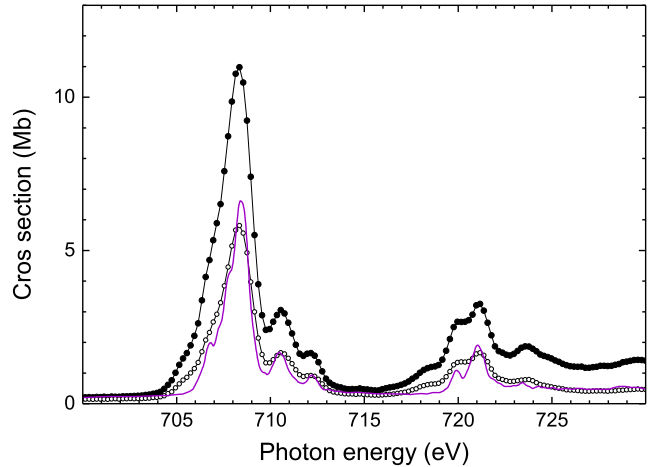}
\caption{\label{fig:comp}Comparison of the present data (full circles: $\sigma_\Sigma$, empty circles: $\sigma_1+\sigma_2$) with the relative ion yields measured by \citet[][full purple line]{Hirsch2012}. The latter have been shifted by +0.6~eV on the photon energy axis and been multiplied by a constant factor to fit to the present cross-section scale.}
\end{figure}

In Fig.~\ref{fig:comp}, the present Fe$^+$ absorption cross section is compared with the result of \citet{Hirsch2012} who used an ion-trap technique in combination with a time-of-flight detection of the product ions. This technique, in principle, facilitates the production of pure ground-state ion-targets \citep{Thissen2008a} but does not allow for the measurement of \textit{absolute} cross sections. The present cross sections were measured on an absolute scale, but with a primary ion beam that comprised ground-state and metastable ions. In this respect, both techniques are complementary. In the Fe$^+$ experiment of \citet{Hirsch2012}, the preparation of pure-ground state targets was aimed at by storing the ions 1--2~s in the ion trap before photoionizing them. However, this storage time may have been somewhat short in view of the recent atomic-structure calculations of \citet{Bautista2015} who obtained much longer lifetimes of up to several hundred seconds for some of the Fe$^+$ metastable levels.

\begin{figure*}
\includegraphics[width=\textwidth]{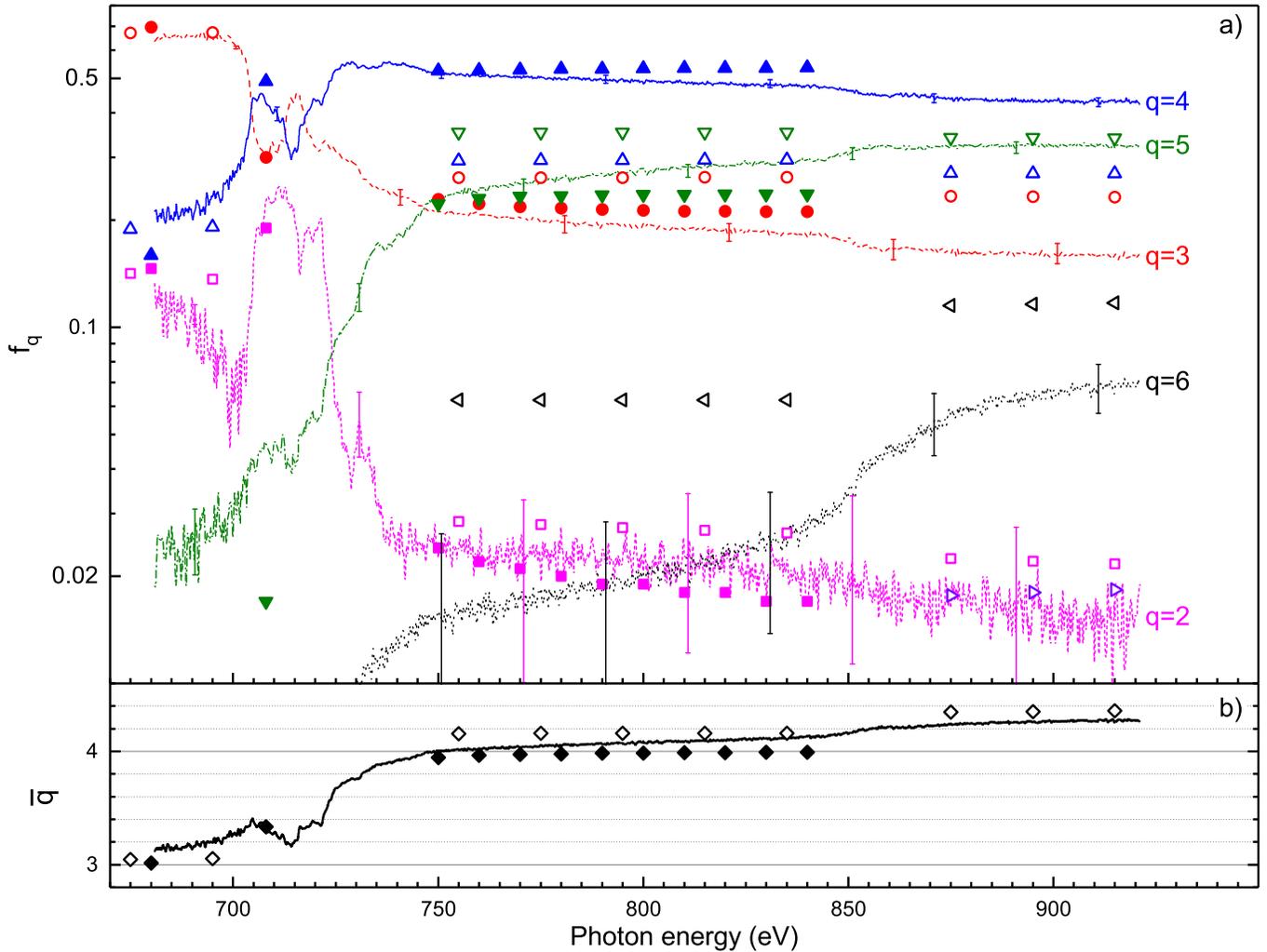}
\caption{\label{fig:fractions}a) Experimental (lines) and theoretical (symbols) results for the product charge-state fractions $f_q$. Data are shown for $q=2$ (magenta short-dashed line \& squares), $q=3$ (red long-dashed line \& circles), $q=4$ (blue full lines \& up-pointing triangles), $q=5$ (olive dash-dotted line \& down-pointing triangles), $q=6$ (black dotted line \& left-pointing triangles), and $q=7$ (purple right-pointing triangles). Not shown are the experimental results for $f_7$ which are too small to be visible on the scale of the figure. Closed symbols denote the results of the present cascade calculations and open symbols denote the results of \citet{Kaastra1993} weighted with the subshell specific photoionization cross sections of \citet{Verner1993a}. Experimental uncertainties are shown by the error bars. b) Mean product charge state $\overline{q}$ from Eq.~\ref{eq:qmean} as a function of the photon energy. The experimental error bars (not shown, see Tab.~\ref{tab:Xsec}) are negligible on the scale of the figure. The closed diamonds represent the results of the present cascade calculations and the open diamonds the results from the cascade calculations of \citet{Kaastra1993} in combination with the cross-section calculations of \citet{Verner1993a}.}
\end{figure*}

In the trap experiment, only doubly and triply charged product ions were detected. Therefore, we compare the relative combined Fe$^{2+}$ and Fe$^{3+}$ ion yield of \citet{Hirsch2012} (full line in Fig.~\ref{fig:comp}) to our summed $\sigma_1+\sigma_2$ (open circles in Fig.~\ref{fig:comp}), which has a dependence on the photon energy different from $\sigma_\Sigma$ (Eq.~\ref{eq:sigmasum}, closed circles in Fig.~\ref{fig:comp}). For this comparison, the relative ion yield of Hirsch e al.\ was multiplied by a constant factor and shifted by 0.6~eV to align resonance positions. This shift is larger than the combined uncertainties ($\sim$0.4~eV) of both experiments. The remaining discrepancy might be explained by different amounts of metastable ions in the two experiments. Apart from the energy shift, there are only minor differences between the two experimental results. In particular, there is agreement with respect to the relative resonance strengths, although the resonance features are narrower in the measurement of Hirsch et al.\ than in the present experiment. This is partly due to their smaller experimental photon energy spread of 125~meV vs.\ 1~eV here and possibly also due to a smaller fraction of metastable primary ions in the trap experiment as compared to the present beam experiment. The otherwise good agreement between both results, as well as our theoretical findings (cf., Fig.~\ref{fig:theo}), strongly suggests that our experimental cross sections are not decisively influenced by the likely presence of metastable ions in the primary Fe$^+$ ion beam.

\subsection{Product-charge state distributions}

\begin{deluxetable*}{lllllllllllll}
\tablecaption{\label{tab:prodfrac}Calculated fractions $F_{k,q}$ of product charge states $q$  after creation of an inner shell hole in subshell $k$ of an Fe$^+$ ion by direct photoionization. Cascades following $2s$ ionization have not been considered in the present calculations which are confined to product charges states $q\leq5$ for reasons given in Sec.~\ref{sec:mcdf}.}
\tablehead{
    \colhead{} &
    \multicolumn{7}{l}{\citet{Kaastra1993}}  &
    \colhead{} &
    \multicolumn{4}{l}{present MCDF theory}\\
    \cline{2-8}\cline{10-13}
    \colhead{Subshell} &
    \colhead{$q$=2}&
    \colhead{3}&
    \colhead{4}&
    \colhead{5}&
    \colhead{6}&
    \colhead{7}&
    \colhead{8}&
    \colhead{}&
    \colhead{$q$=2}&
    \colhead{3}&
    \colhead{4}&
    \colhead{5}
    }
\startdata
 $2s_{1/2}$  &        & 0.0170 & 0.0953 & 0.2581 & 0.4829 & 0.1409 & 0.0058   &   &       &        &         &        \\
 $2p_{1/2}$  & 0.0158 & 0.2606 & 0.3076 & 0.3542 & 0.0617 & 0.0001 &          &   &       & 0.197  & 0.421   &  0.382  \\
 $2p_{3/2}$  & 0.0153 & 0.1770 & 0.3061 & 0.4256 & 0.0759 & 0.0001 &          &   &       & 0.170  & 0.507   &  0.323  \\
 $3s$        &    	  & 0.0437 & 0.9563 &        &        &    	   &          &   &       & 0.159  & 0.841   &	     \\
 $3p$        & 0.0001 & 0.9999 &        &	     &        & 	   &          &   &       & 1.0    &         &	     \\
 $3d$        & 1.0    &        &        &	     &        &    	   &          &   &  1.0  &        &         &        \\
\enddata
\end{deluxetable*}

\begin{deluxetable}{llllll}
\tablecaption{\label{tab:relxsec}Calculated relative cross sections $\tilde{\sigma}_k/\sigma_\mathrm{tot}$ for subshell specific photoionization of Fe$^+$ ions. Values are shown for two typical energies where resonant processes do not play a role.}
\tablehead{
    \colhead{Energy} &
    \multicolumn{5}{c}{$\tilde{\sigma}_k/\sigma_\mathrm{tot}$}\\
    \cline{2-6}
    \colhead{(eV)}&
    \colhead{$2s$}&
    \colhead{$2p$}&
    \colhead{$3s$}&
    \colhead{$3p$}&
    \colhead{$3d$}
    }
\startdata
 & \multicolumn{5}{l}{\citet{Verner1993a}}\\
 680 &       &       & 0.20  & 0.66   &	0.14\\
 882 & 0.13  & 0.77  & 0.024 &	0.073 & 0.010\\
\hline
 & \multicolumn{5}{l}{present MCDF theory}\\
 680 &       &        &  0.19  & 0.66  &  0.15\\
 882 & 0.096 &	0.77  &  0.03  & 0.089 &  0.015\\
\enddata
\end{deluxetable}

Since we have measured the cross sections for single to sixfold ionization of Fe$^+$ on an absolute scale, the product charge-state fractions as a function of photon energy  are readily derived as $f_q=\sigma_{q-1}/\sigma_\Sigma$,  with $\sigma_\Sigma$ from Eq.~\ref{eq:sigmasum}. These fractions are shown in Fig.~\ref{fig:fractions}a. It should be noted that the systematic uncertainty of the cross-section scale cancels out for the $f_q$ values. The mean product charge-state can then be calculated as
\begin{equation}
\label{eq:qmean} \overline{q} = \sum_{q=2}^7 q f_q = \frac{1}{\sigma_\Sigma}\sum_{m=1}^6(m+1)\sigma_m.
\end{equation}
The experimental mean product charge-state is shown in Fig.~\ref{fig:fractions}b. It rises from 3.1 at the lowest experimental photon energy to about 4 at energies above the $2p$-ionization threshold and further increases to 4.3 above the $2s$-ionization threshold (Tab.~\ref{tab:Xsec}).

\begin{figure}
\includegraphics[width=\figurewidth]{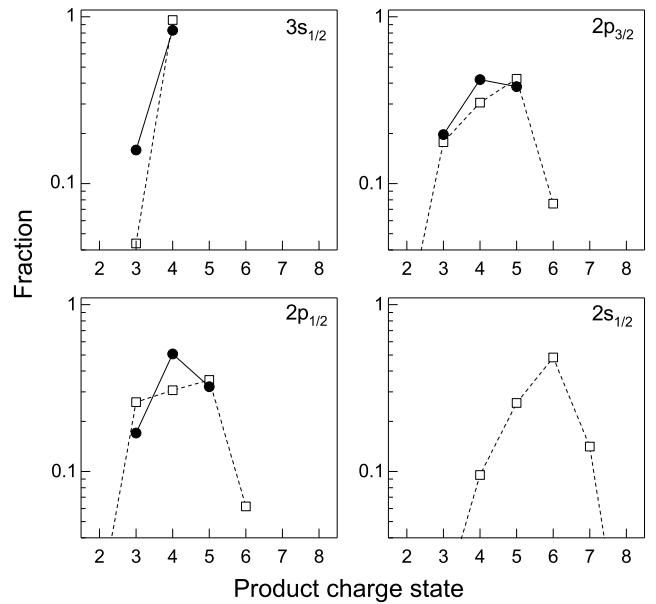}
\caption{\label{fig:prodfrac} Fractions $F_{k,q}$ produced by direct photoionization of the atomic subshell $k$ resulting in product charge states $q$ (from Tab.~\ref{tab:prodfrac}). Present theoretical results (full circles) and results of \citet[][open squares]{Kaastra1993}. The lines are drawn to guide the eye.}
\end{figure}

As discussed above, the theoretical calculation of the product charge-state fractions is a formidable challenge since it involves keeping track of cascades of  autoionizing and radiative transitions via all energetically accessible intermediate levels. Employing the configuration-average approximation, \citet{Kaastra1993} have carried out simplified cascade calculations for astrophysically relevant ions including iron ions. They arrived at fractions $F_{k,q}$ of product charge states $q$ produced by direct photoionization of the atomic subshell $k$. Their results for Fe$^+$ are listed in Tab.~\ref{tab:prodfrac} and are plotted in Fig.~\ref{fig:prodfrac} as open squares. For comparison with the experimental fractions $f_q$, the fractions $F_{k,q}$ have to be weight\-ed with the probabilities for the creation of a hole in subshell $k$ and then summed over all relevant subshells, i.e.,
\begin{equation}\label{eq:fq}
f_q = \frac{1}{\sigma_\mathrm{tot}(E_\mathrm{ph})}\sum_k \tilde{\sigma}_k(E_\mathrm{ph})F_{k,q}
\end{equation}
where $\tilde{\sigma}_k(E_\mathrm{ph})$ is the cross section for direct photoionization of subshell $k$, which depends on the photon energy $E_\mathrm{ph}$, and $\sigma_\mathrm{tot}(E_\mathrm{ph}) = \sum_k \tilde{\sigma}_k(E_\mathrm{ph})$ is the total cross section. For the evaluation of Eq.~\ref{eq:fq}, we have used the fractions $F_{k,q}$ of \citet[][given here in Tab.~\ref{tab:prodfrac}]{Kaastra1993} and the subshell-specific cross sections $\tilde{\sigma}_k(E_\mathrm{ph})$ from \citet[][given here in Tab.~\ref{tab:relxsec}]{Verner1993a}. As discussed above, the corresponding total cross section (dashed line in Fig.~\ref{fig:absorp}) agrees reasonably well with the experimental absorption cross section in the photon-energy ranges without photoionization resonances. The resulting theoretical fractions $f_q$ and mean charge state are plotted in Fig.~\ref{fig:fractions} using open symbols. The calculated mean charge state closely follows the experimental trend. The maximum discrepancy between theory and experiment is a difference in mean charge state of 0.14. The discrepancies are much larger when individual charge-state fractions are considered. For example, the fractions $f_3$ and $f_5$ are overestimated by up to $\sim$45\% and $f_4$ is underestimated by up to $\sim$40\%.

Compared to the work of \citet{Kaastra1993}, the present cascade model and cross-section calculations result in better agreement with the measured charge state distributions (Fig.~\ref{fig:fractions}a). This is due to our construction of a decay tree that takes fine-structure levels into account (Sec.~\ref{sec:mcdf}). At energies where the photoionization cross section does not exhibit thresholds or resonances, our relative cross sections for subshell specific photo\-ionization are very similar to the results of \citet[][Tab.~\ref{tab:relxsec}]{Verner1993a}. However, our fractions $F_{k,q}$ are significantly different from those of \citet{Kaastra1993}. Both sets of results are listed in Table~\ref{tab:prodfrac} and displayed in Figure \ref{fig:prodfrac}. For the purpose of this comparison, statistical averages of our fine-structure resolved results are presented. Cascades resulting from a $2s$ vacancy have not been accounted for in our MCDF calculations because their treatment on the fine-structure level has been found to be too complex to be manageable within the scope of the present calculations.

As can be seen in Fig.~\ref{fig:fractions}, the present cascade calculations predict a significantly larger fraction of fourfold charged product ions than what results from  the work of \citet{Kaastra1993}. As a consequence, the present ion yields compare well to the experimental results. Notably, the cross\-over between Fe$^{3+}$ and Fe$^{5+}$ at around 750~eV is reproduced. Nevertheless, there are still slight discrepancies with the experimental findings. For higher energies, the computed ion yields remain almost constant, where\-as in the experiment a significant shift towards higher charge states is observed. For increasing photon energies between 750~eV and 850~eV the production of Fe$^{5+}$ increases while both Fe$^{3+}$ and Fe$^{4+}$ de\-crease. The decrease of Fe$^{3+}$ is reproduced in the computations, however Fe$^{4+}$ is predicted to also increase, and as a consequence, the production of Fe$^{5+}$ is slightly underestimated for higher photon energies.

Higher charge states are significantly underestimated or absent in our computations because multiple Auger processes \citep{Mueller2015a} and shake-up processes \citep[see, e.g.,][]{Schippers2016a} are neglected. \changed{Direct multiple Auger processes and shake-up transitions are expected to mainly contribute to the very high charge states such as $q=6, 7$. Furthermore shake-down processes play an important role in the decay of double-core hole vacancies in higher charge states that are too low in energy to decay by a two-electron Auger process but are still above the ionization threshold. In such a situation, a rather slow Auger decay with an additional shake-down becomes important, and these processes are still expected to be fast compared to radiative decays.} Similarly, the photo\-ion\-ization can also be accompanied by direct double ionization which becomes important for photon energies further away from the ion\-ization threshold. Neglecting these processes may be a possible cause for the remaining discrepancies of the computed ion yields compared to the measurements for high photon energies.

\section{Summary and conclusions}\label{sec:sum}

Building upon our recent improvements of the photon-ion merged-beams technique, absolute cross sections for single and multiple (up to sixfold) ionization of Fe$^+$ following $L$-shell excitation and ionization have been measured with unprecedented sensitivity. Strong ionization resonances due to $2p\to nd$ excitations with $n=3,4$ and 5 are visible in all partial cross sections. A complication of the present experimental technique arises from the presence of metastable ions in the ion beam. However, the comparison of the present experimental cross sec\-tions with the ion-yield of an ion-trap experiment, where fewer metastable ions were expected, suggests that our measured cross sections are not significantly influenced by the likely presence of  metastable Fe$^+$ ions.

The sum of all our measured $m$-fold photoionization cross sections agrees well with our computed absorption cross section which includes resonant ionization processes. In the energy ranges where no resonance features are present, the cross section of \citet{Verner1993a} for direct photoionization of Fe$^+$ agrees with our experimental results to within the experimental uncertainty, except at energies above the $2s$ ionization threshold where direct double (and multiple) ionization might also play a role. In addition, we have computationally tracked the deexcitation cascades that set in after the initial creation of an inner-shell hole and that leads to distributions of product charge states. Compared to earlier calculations by \citet{Kaastra1993}, who used the configuration-average approximation, the present cascade calculations have been carried out on the fine-structure level, and are in much better agreement with the experimental findings.

Since the experimental energy scale has been determined to within 0.2~eV, the present results can be used to identify gas-phase Fe$^+$ $L$-shell features in astronomical observations and, thereby, contribute to an improved determination of the iron abundance in the ISM. Moreover, our absolute cross sections, which have been determined with a $\pm$15\% systematic uncertainty (at a 90\% confidence level), are valuable for opacity calculations for matter exposed to soft X-rays. The analysis of similar experimental data for Fe$^{2+}$ and Fe$^{3+}$ ions is under way and the corresponding results will be presented in forthcoming publications.

\begin{acknowledgments}
This research was carried out at the light source PETRA\,III at DESY, a member of the Helmholtz Association (HGF). We would like to thank G.~Hartmann, F.~Scholz, and J.~Seltmann, for assistance in using beamline P04. This research has been funded in part by the German Federal Ministry for Education and Research (BMBF) within the \lq\lq{}Verbundforschung\rq\rq\ funding scheme under contracts 05K16GUC, 05K16RG1, and 05K16SJA. S.B. and K.S. would like to thank SFB 755, \lq\lq{}Nanoscale photonic imaging, project B03\rq\rq\ for financial support. S.B. acknowledges funding from the Initiative and Networking Fund of the Helmholtz Association. D.W.S. was supported, in part, by the NASA Astrophysics Research and Analysis program.
\end{acknowledgments}

\appendix

\section{Contributions by sequential two-photon ionization}

\changed{

Assuming a uniform photon flux density, the count rate of $q$-times charged product ions resulting from one-photon ionization of $p$-times charged primary ions can be calculated from the cross section $\sigma_{p\to q}$ for $m$-fold ionization ($m=q-p$) as
\begin{equation}\label{eq:r1}
R_q^{(1)} = \frac{N_\mathrm{ph}}{A\tau}N_p\sigma_{p\to q},
\end{equation}
where $N_p$ is the number of primary ions in the photon-ion interaction region and ${N_\mathrm{ph}}$  the number of photons passing through a suitably defined effective target area $A$ during the time interval $\tau$ that an ion spends in the interaction region. During this time interval, an ion may be further ionized by a second photon. The corresponding two-photon ionization rate is
\begin{equation}\label{eq:r2a}
R_q^{(2)} = \frac{1}{2}\frac{N_\mathrm{ph}}{A\tau}\sum_{q'=p+1}^{q-1}N_{q'}\sigma_{q'\to q}
\end{equation}
where the summation is over all intermediate charge states $q'$ with $p<q'<q$. The factor $1/2$ accounts for the fact that the $q'$-times charged ions, on the average, are only available on half the interaction length. Inserting $N_{q'} = R_{q'}^{(1)}\tau$ with $R_{q'}^{(1)}$ from Eq.~\ref{eq:r1} gives
\begin{equation}\label{eq:r2}
R_q^{(2)} = \frac{1}{2}N_p\tau\left(\frac{N_\mathrm{ph}}{A\tau}\right)^2 \sum_{q'=p+1}^{q-1}\sigma_{p\to q'}\sigma_{q'\to q}.
\end{equation}
Thus, the ratio $\kappa$ of count rates due to two-photon and one-photon ionization is
\begin{equation}\label{eq:rr}
\kappa = \frac{R_q^{(2)}}{R_q^{(1)}} = \frac{N_\mathrm{ph}}{2A}\sum_{q'=p+1}^{q-1}\frac{\sigma_{p\to q'}\sigma_{q'\to q}}{\sigma_{p\to q}}.
\end{equation}

Under the assumption that the probabilities for two-photon ionization are small, the cross sections $\sigma_{p\to q'}$ in Eq.~\ref{eq:rr} may be substituted by the measured cross sections $\sigma_m^{(p)} = \sigma_{p\to(p+m)}$ for $m$-fold ionization of $p$-fold charged primary ions with $m=q'-p$. Since the cross sections $\sigma_{q'\to q} = \sigma_{q'\to q'+(q-q')}$ for the photoionization of ions with primary charge states $q'>p$ are not known, and since we need only an upper limit for the ratio $\kappa$ we replace $\sigma_{q'\to q}$ by (the most likely bigger) $\sigma_{p\to p +(q-q')} = \sigma_k^{(p)}$  with $k = q-q'$ and obtain
\begin{equation}\label{eq:rm}
\kappa \lesssim \frac{N_\mathrm{ph}}{2A}\sum_{q'=p+1}^{q-1}\frac{\sigma_{q'-p}^{(p)}\sigma_{q-q'}^{(p)}}{\sigma_{q-p}^{(p)}}.
\end{equation}
In the present case $p=1$. The nomenclature introduced in the main text requires that the cross sections for $m$-fold ionization of singly charged iron ions is $\sigma_m^{(1)} = \sigma_m$. Hence, in this case
\begin{equation}\label{eq:kappa1}
\kappa \lesssim \frac{N_\mathrm{ph}}{2A} \frac{1}{\sigma_{q-p}} \sum_{q'=2}^{q-1}\sigma_{q'-1}\sigma_{q-q'}.
\end{equation}

For the calculation of $N_\mathrm{ph}$ one should consider that the photon beam from a synchrotron has a pulse structure. At the PETRA\,III synchrotron, the photon pulse duration is $\Delta t=44$~ps. In the present case, the time difference between two pulses was up to 16~ns. This is much smaller than the ion flight time $\tau = 3.4$~$\mu$s through the interaction region. Under this condition
$N_\mathrm{ph} = \phi_\mathrm{ph} \tau$ where $\phi_\mathrm{ph}$ denotes the time-averaged photon flux. In the present experiment it was at most $\phi=4\times10^{13}$~s$^{-1}$ (Sec.~\ref{sec:exp}). The effective target area was determined by beam-overlap measurements to $A=2.1$~mm$^2$ \citep[for details see][]{Schippers2014}. With the given numerical values one obtains $N_\mathrm{ph}/(2A) = \phi_\mathrm{ph}\tau/(2A) = 3.2\times 10^{-9}$~Mb$^{-1}$.

\begin{figure}
\centering{\includegraphics[width=0.5\textwidth]{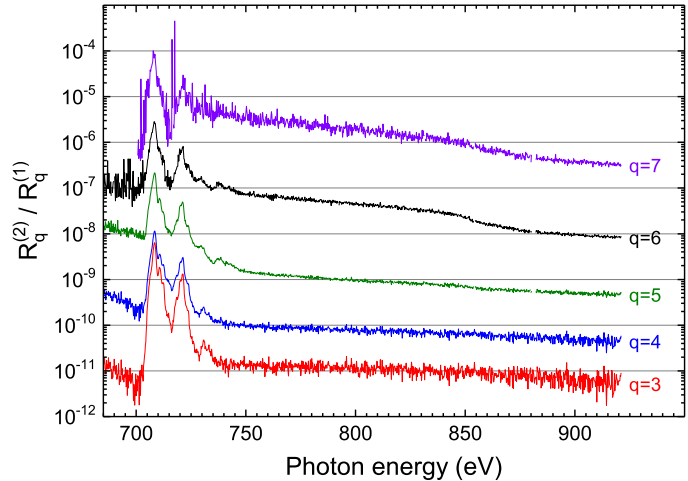}}
\caption{\label{fig:rr}\changed{Ratios of two-photon vs.\ one-photon photoionization count rates for product charge states $q>2$. The color coding is the same as in Fig.~\ref{fig:fractions}. The curves were calculated using $N_\mathrm{ph}/(2A) = 3.2\times 10^{-9}$~Mb$^{-1}$, $p=1$, and the measured cross sections from Fig.~\ref{fig:all} in Eq.~\ref{eq:rm} (see text).}}
\end{figure}

The calculated ratios $R_q^{(2)}/R_q^{(1)}$ are displayed in Fig.~\ref{fig:rr} as functions of photon energy for all measured product charge states. It should be noted, that the curves overestimate the two-photon contribution, in particular, in the energy regions which are dominated by photoionization resonances because of the substitution of the cross sections for $m$-fold ionization of Fe$^{p+}$ ions with $p>1$ by the corresponding Fe$^+$ cross sections. For more highly charged primary ions the photoionization resonances can be expected to shift to other energies. This should considerably flatten out the resonance structures in Fig.~\ref{fig:rr}. From this figure we finally conclude that sequential two-photon ionization is negligible in all product-ion channels and at all experimental photon energies.}


\begin{thebibliography}{}
\expandafter\ifx\csname natexlab\endcsname\relax\def\natexlab#1{#1}\fi
\providecommand{\url}[1]{\href{#1}{#1}}

\bibitem[{Andersson {et~al.}(2015)Andersson, Beerwerth, Linusson, Eland,
  Zhaunerchyk, Fritzsche, \& Feifel}]{Andersson2015}
Andersson, J., Beerwerth, R., Linusson, P., {et~al.} 2015, PhRvA, 92, 023414.
\newblock \url{http://link.aps.org/doi/10.1103/PhysRevA.92.023414}

\bibitem[{{Badnell} {et~al.}(2005){Badnell}, {Bautista}, {Butler}, {Delahaye},
  {Mendoza}, {Palmeri}, {Zeippen}, \& {Seaton}}]{Badnell2005a}
{Badnell}, N.~R., {Bautista}, M.~A., {Butler}, K., {et~al.} 2005, MNRAS, 360,
  458.
\newblock \url{http://dx.doi.org/10.1111/j.1365-2966.2005.08991.x}

\bibitem[{Badnell \& Seaton(2003)}]{Badnell2003b}
Badnell, N.~R., \& Seaton, M.~J. 2003, JPhB, 36, 4367.
\newblock \url{https://doi.org/10.1088/0953-4075/36/21/015}

\bibitem[{{Bautista} {et~al.}(2015){Bautista}, {Fivet}, {Ballance}, {Quinet},
  {Ferland}, {Mendoza}, \& {Kallman}}]{Bautista2015}
{Bautista}, M.~A., {Fivet}, V., {Ballance}, C., {et~al.} 2015, ApJ, 808, 174.
\newblock \url{http://dx.doi.org/10.1088/0004-637X/808/2/174}

\bibitem[{Beiersdorfer {et~al.}(2013)Beiersdorfer, Lepson, D\'iaz, Ishikawa, \&
  Tr\"{a}bert}]{Beiersdorfer2013}
Beiersdorfer, P., Lepson, J.~K., D\'iaz, F., Ishikawa, Y., \& Tr\"{a}bert, E.
  2013, Phys. Scr., T156, 014007.
\newblock \url{https://doi.org/10.1088/0031-8949/2013/T156/014007}

\bibitem[{Bernhardt {et~al.}(2014)Bernhardt, Becker, Grieser, Hahn, Krantz,
  Lestinsky, Novotn\'y, Repnow, Savin, Spruck, Wolf, M\"uller, \&
  Schippers}]{Bernhardt2014}
Bernhardt, D., Becker, A., Grieser, M., {et~al.} 2014, PhRvA, 90, 012702.
\newblock \url{http://dx.doi.org/10.1103/PhysRevA.90.012702}

\bibitem[{Bernitt {et~al.}(2012)Bernitt, Brown, Rudolph, Steinbr{\"u}gge, Graf,
  Leutenegger, Epp, Eberle, Kubi\v{c}ek, M{\"a}ckel, Simon, Tr{\"a}bert, Magee,
  Beilmann, Hell, Schippers, M{\"u}ller, Kahn, Surzhykov, Harman, Keitel,
  Clementson, Porter, Schlotter, Turner, Ullrich, Beiersdorfer, \& {Crespo
  L{\'o}pez-Urrutia}}]{Bernitt2012}
Bernitt, S., Brown, G.~V., Rudolph, J.~K., {et~al.} 2012, Natur, 492, 225.
\newblock \url{http://dx.doi.org/10.1038/nature11627}

\bibitem[{Bizau {et~al.}(2006)Bizau, Blancard, Cubaynes, Folkmann, Kilbane,
  Faussurier, Luna, Lemaire, Blieck, \& Wuilleumier}]{Bizau2006a}
Bizau, J.~M., Blancard, C., Cubaynes, D., {et~al.} 2006, PhRvA, 73, 020707.
\newblock \url{http//dx.doi.org/10.1103/PhysRevA.73.020707}

\bibitem[{Buth {et~al.}(2017)Buth, Beerwerth, Obaid, Berrah, Cederbaum, \&
  Fritzsche}]{Buth2017}
Buth, C., Beerwerth, R., Obaid, R., {et~al.} 2017, arXiv:1705.07521.
\newblock \url{https://arxiv.org/abs/1705.07521}

\bibitem[{Chen {et~al.}(2006)Chen, Gu, Beiersdorfer, Boyce, Brown, Kahn,
  Kelley, Kilbourne, Porter, \& Scofield}]{Chen2006a}
Chen, H., Gu, M.~F., Beiersdorfer, P., {et~al.} 2006, ApJ, 646, 653.
\newblock \url{http://dx.doi.org/10.1086/504708}

\bibitem[{Cowan(1981)}]{Cowan1981}
Cowan, R.~D. 1981, The Theory of Atomic Structure and Spectra (Berkeley:
  California University Press), atomic-structure code accessible via
  \url{http://aphysics2.lanl.gov/tempweb/lanl/}

\bibitem[{Dumitriu {et~al.}(2010)Dumitriu, Bilodeau, Gorczyca, Walter, Gibson,
  Aguilar, Pe\ifmmode \check{s}\else \v{s}\fi{}i\ifmmode~\acute{c}\else
  \'{c}\fi{}, Rolles, \& Berrah}]{Dumitriu2010}
Dumitriu, I., Bilodeau, R.~C., Gorczyca, T.~W., {et~al.} 2010, PhRvA, 053404.
\newblock \url{https://doi.org/10.1103/PhysRevA.81.053404}

\bibitem[{{El Hassan} {et~al.}(2009){El Hassan}, Bizau, Blancard, Cosse,
  Cubaynes, Faussurier, \& Folkmann}]{ElHassan2009}
{El Hassan}, N., Bizau, J.~M., Blancard, C., {et~al.} 2009, PhRvA, 79, 033415.
\newblock \url{https://doi.org/10.1103/PhysRevA.79.033415}

\bibitem[{Epp {et~al.}(2007)Epp, {Crespo L\'opez-Urrutia}, Brenner, M\"ackel,
  Mokler, Treusch, Kuhlmann, Yurkov, Feldhaus, Schneider, Wellh{\"o}fer,
  Martins, Wurth, \& Ullrich}]{Epp2007}
Epp, S.~W., {Crespo L\'opez-Urrutia}, J.~R., Brenner, G., {et~al.} 2007, PhRvL,
  98, 183001.
\newblock \url{https://doi.org/10.1103/PhysRevLett.98.183001}

\bibitem[{Fabian {et~al.}(2009)Fabian, Zoghbi, Ross, Uttley, Gallo, Brandt,
  Blustin, Boller, Caballero-Garcia, Larsson, Miller, Miniutti, Ponti, Reis,
  Reynolds, Tanaka, \& Young}]{Fabian2009}
Fabian, A.~C., Zoghbi, A., Ross, R.~R., {et~al.} 2009, Nature, 459, 540.
\newblock \url{http://dx.doi.org/10.1038/nature08007}

\bibitem[{Fano(1961)}]{Fano1961}
Fano, U. 1961, PhRv, 124, 1866.
\newblock \url{https://doi.org/10.1103/PhysRev.124.1866}

\bibitem[{Fano \& Cooper(1965)}]{Fano1965}
Fano, U., \& Cooper, J.~W. 1965, PhRv, 137, A1364.
\newblock \url{https://doi.org/10.1103/PhysRev.137.A1364}

\bibitem[{Fivet {et~al.}(2012)Fivet, Bautista, \& Ballance}]{Fivet2012}
Fivet, V., Bautista, M.~A., \& Ballance, C.~P. 2012, JPhB, 45, 035201.
\newblock \url{https://doi.org/10.1088/0953-4075/45/3/035201}

\bibitem[{Fritzsche(2012)}]{Fritzsche2012a}
Fritzsche, S. 2012, CPC, 183, 1525 .
\newblock \url{https://doi.org/10.1016/j.cpc.2012.02.016}

\bibitem[{Gatuzz {et~al.}(2015)Gatuzz, Garcia, Kallman, Mendoza, \&
  Gorczyca}]{Gatuzz2015}
Gatuzz, E., Garcia, J., Kallman, T.~R., Mendoza, C., \& Gorczyca, T.~W. 2015,
  ApJ, 800, 29.
\newblock \url{http://dx.doi.org/10.1088/0004-637X/800/1/29}

\bibitem[{{Gatuzz} {et~al.}(2016){Gatuzz}, {Garc{\'{\i}}a}, {Kallman}, \&
  {Mendoza}}]{Gatuzz2016}
{Gatuzz}, E., {Garc{\'{\i}}a}, J.~A., {Kallman}, T.~R., \& {Mendoza}, C. 2016,
  A\&A, 588, A111.
\newblock \url{http://dx.doi.org/10.1051/0004-6361/201527752}

\bibitem[{Gharaibeh {et~al.}(2011)Gharaibeh, Aguilar, Covington, Emmons,
  Scully, Phaneuf, M\"uller, Bozek, Kilcoyne, Schlachter, \'Alvarez, Cisneros,
  \& Hinojosa}]{Gharaibeh2011}
Gharaibeh, M.~F., Aguilar, A., Covington, A.~M., {et~al.} 2011, PhRvA, 83,
  043412.
\newblock \url{http://dx.doi.org/10.1103/PhysRevA.83.043412}

\bibitem[{Gu {et~al.}(2001)Gu, Kahn, Savin, Behar, Beiersdorfer, Brown,
  Liedahl, \& Reed}]{Gu2001}
Gu, M.~F., Kahn, S.~M., Savin, D.~W., {et~al.} 2001, ApJ, 563, 462.
\newblock \url{https://doi.org/10.1086/323683}

\bibitem[{{Hahn} {et~al.}(2015){Hahn}, {Becker}, {Bernhardt}, {Grieser},
  {Krantz}, {Lestinsky}, {M{\"u}ller}, {Novotn{\'y}}, {Repnow}, {Schippers},
  {Spruck}, {Wolf}, \& {Savin}}]{Hahn2015}
{Hahn}, M., {Becker}, A., {Bernhardt}, D., {et~al.} 2015, ApJ, 813, 16.
\newblock \url{http://dx.doi.org/10.1088/0004-637X/813/1/16/pdf}

\bibitem[{Hansen {et~al.}(2007)Hansen, Kjeldsen, Folkmann, Martins, \&
  West}]{Hansen2007a}
Hansen, J.~E., Kjeldsen, H., Folkmann, F., Martins, M., \& West, J.~B. 2007,
  JPhB, 40, 293.
\newblock \url{http://dx.doi.org/10.1088/0953-4075/40/2/005}

\bibitem[{Hirsch {et~al.}(2012)Hirsch, Zamudio-Bayer, Ameseder, Langenberg,
  Rittmann, Vogel, M\"oller, {v. Issendorff}, \& Lau}]{Hirsch2012}
Hirsch, K., Zamudio-Bayer, V., Ameseder, F., {et~al.} 2012, PhRvA, 85, 062501.
\newblock \url{http://dx.doi.org/10.1103/PhysRevA.85.062501}

\bibitem[{{Hummer} {et~al.}(1993){Hummer}, {Berrington}, {Eissner}, {Pradhan},
  {Saraph}, \& {Tully}}]{Hummer1993}
{Hummer}, D.~G., {Berrington}, K.~A., {Eissner}, W., {et~al.} 1993, A\&A, 279,
  298.
\newblock \url{http://adsabs.harvard.edu/abs/1993A%26A...279..298H}

\bibitem[{Jenkins(2009)}]{Jenkins2009}
Jenkins, E.~B. 2009, ApJ, 700, 1299.
\newblock \url{http://dx.doi.org/10.1088/0004-637X/700/2/1299}

\bibitem[{{Jensen} \& {Snow}(2007)}]{Jensen2007}
{Jensen}, A.~G., \& {Snow}, T.~P. 2007, ApJ, 669, 378.
\newblock \url{http://dx.doi.org/10.1086/521638}

\bibitem[{J{\"o}nsson \& Andersson(2007)}]{Joensson2007}
J{\"o}nsson, P., \& Andersson, M. 2007, JPhB, 40, 2417.
\newblock \url{https://doi.org/10.1088/0953-4075/40/12/016}

\bibitem[{Juett {et~al.}(2006)Juett, Schulz, Chakrabarty, \&
  Gorczyca}]{Juett2006}
Juett, A.~M., Schulz, N.~S., Chakrabarty, D., \& Gorczyca, T.~W. 2006, ApJ,
  648, 1066.
\newblock \url{https://doi.org/10.1086/506189}

\bibitem[{{Kaastra} \& {Mewe}(1993)}]{Kaastra1993}
{Kaastra}, J.~S., \& {Mewe}, R. 1993, A\&AS, 97, 443.
\newblock \url{http://cdsads.u-strasbg.fr/abs/1993A%26AS...97..443K}

\bibitem[{Kjeldsen(2006)}]{Kjeldsen2006a}
Kjeldsen, H. 2006, JPhB, 39, R325.
\newblock \url{https://doi.org/10.1088/0953-4075/39/21/R01}

\bibitem[{Kjeldsen {et~al.}(2002)Kjeldsen, Kristensen, Folkmann, \&
  Andersen}]{Kjeldsen2002c}
Kjeldsen, H., Kristensen, B., Folkmann, F., \& Andersen, T. 2002, JPhB, 35,
  3655.
\newblock \url{https://doi.org/10.1088/0953-4075/35/17/303}

\bibitem[{Koivisto {et~al.}(1994)Koivisto, \"Arje, \& Nurmia}]{Koivisto1994}
Koivisto, H., \"Arje, J., \& Nurmia, M. 1994, NIMB, 94, 291 .
\newblock \url{http://dx.doi.org/10.1016/0168-583X(94)95368-6}

\bibitem[{Kramida {et~al.}(2015)Kramida, Ralchenko, Reader, \& {NIST ASD
  Team}}]{Kramida2015}
Kramida, A., Ralchenko, Y., Reader, J., \& {NIST ASD Team}. 2015, NIST Atomic
  Spectra Database (version 5.3), [Online]. Available:
  http://physics.nist.gov/asd, Tech. rep., National Institute of Standards and
  Technology.
\newblock \url{http://physics.nist.gov/asd}

\bibitem[{Ku\v{c}as {et~al.}(2015)Ku\v{c}as, Momkauskait\.{e}, \&
  Karazija}]{Kucas2015}
Ku\v{c}as, S., Momkauskait\.{e}, A., \& Karazija, R. 2015, ApJ, 810, 26.
\newblock \url{https://doi.org/10.1088/0004-637X/810/1/26}

\bibitem[{Madden \& Codling(1963)}]{Madden1963a}
Madden, R.~P., \& Codling, K. 1963, PhRvL, 10, 516 .
\newblock \url{https://doi.org/10.1103/PhysRevLett.10.116}

\bibitem[{Martins(2001)}]{Martins2001}
Martins, M. 2001, JPhB, 34, doi:10.1088/0953-4075/34/7/313.
\newblock \url{https://doi.org/10.1088/0953-4075/34/7/313}

\bibitem[{Martins {et~al.}(2006)Martins, Godehusen, Richter, Wernet, \&
  Zimmermann}]{Martins2006a}
Martins, M., Godehusen, K., Richter, T., Wernet, P., \& Zimmermann, P. 2006,
  JPhB, 39, R79.
\newblock \url{http://dx.doi.org/10.1088/0953-4075/39/5/R01}

\bibitem[{McGuire(1972)}]{McGuire1972}
McGuire, E.~J. 1972, PhRvA, 5, 1043.
\newblock \url{http://link.aps.org/doi/10.1103/PhysRevA.5.1043}

\bibitem[{Miedema \& de~Groot(2013)}]{Miedema2013}
Miedema, P.~S., \& de~Groot, F.~M. 2013, JESRP, 187, 32 .
\newblock \url{http://dx.doi.org/10.1016/j.elspec.2013.03.005}

\bibitem[{Mies(1968)}]{Mies1968a}
Mies, F.~H. 1968, PhRv, 175, 164.
\newblock \url{https://doi.org/10.1103/PhysRev.175.164}

\bibitem[{Monte {et~al.}(2006)Monte, Santos, Fule, Fonseca, \&
  Sousa}]{Monte2006}
Monte, M. J.~S., Santos, L. M. N. B.~F., Fulem, M., Fonseca, J. M.~S., \& Sousa,
  C. A.~D. 2006, J. Chem. Eng. Data, 51, 757.
\newblock \url{http://dx.doi.org/10.1021/je050502y}

\bibitem[{M\"uller {et~al.}(2015{\natexlab{a}})M\"uller, Schippers, Hellhund,
  Holste, Kilcoyne, Phaneuf, Ballance, \& McLaughlin}]{Mueller2015c}
M\"uller, A., Schippers, S., Hellhund, J., {et~al.} 2015{\natexlab{a}}, JPhB,
  48, 235203.
\newblock \url{http://dx.doi.org/10.1088/0953-4075/48/23/235203}

\bibitem[{M{\"u}ller {et~al.}(2009)M{\"u}ller, Schippers, Phaneuf, Scully,
  Aguilar, Covington, {\'A}lvarez, Cisneros, Emmons, Gharaibeh, Hinojosa,
  Schlachter, \& McLaughlin}]{Mueller2009a}
M{\"u}ller, A., Schippers, S., Phaneuf, R.~A., {et~al.} 2009, JPhB, 42, 235602.
\newblock \url{http://dx.doi.org/10.1088/0953-4075/42/23/235602}

\bibitem[{M\"{u}ller {et~al.}(2014)M\"{u}ller, Schippers, Phaneuf, Scully,
  Aguilar, Cisneros, Gharaibeh, Schlachter, \& McLaughlin}]{Mueller2014a}
M\"{u}ller, A., Schippers, S., Phaneuf, R.~A., {et~al.} 2014, JPhB, 47, 135201.
\newblock \url{https://doi.org/10.1088/0953-4075/47/13/135201}

\bibitem[{M\"uller {et~al.}(2015{\natexlab{b}})M\"uller, {Borovik, Jr.}, Buhr,
  Hellhund, Holste, Kilcoyne, Klumpp, Martins, Ricz, Viefhaus, \&
  Schippers}]{Mueller2015a}
M\"uller, A., {Borovik, Jr.}, A., Buhr, T., {et~al.} 2015{\natexlab{b}}, PhRvL,
  114, 013002.
\newblock \url{http://dx.doi.org/10.1103/PhysRevLett.114.013002}

\bibitem[{M\"uller {et~al.}(2017)M\"uller, Bernhardt, {Borovik, Jr.}, Buhr,
  Hellhund, Holste, Kilcoyne, Klumpp, Martins, Ricz, Seltmann, Viefhaus, \&
  Schippers}]{Mueller2017}
M\"uller, A., Bernhardt, D., {Borovik, Jr.}, A., {et~al.} 2017, ApJ, 836, 166.
\newblock \url{https://doi.org/10.3847/1538-4357/836/2/166}

\bibitem[{Nahar(1997)}]{Nahar1997a}
Nahar, S.~N. 1997, PhRvA, 55, 1980.
\newblock \url{http://dx.doi.org/10.1103/PhysRevA.55.1980}

\bibitem[{Nahar \& Pradhan(1994)}]{Nahar1994}
Nahar, S.~N., \& Pradhan, A.~K. 1994, JPhB, 27, 429.
\newblock \url{https://doi.org/10.1088/0953-4075/27/3/010}

\bibitem[{Richter {et~al.}(2004)Richter, Godehusen, Martins, Wolff, \&
  Zimmermann}]{Richter2004}
Richter, T., Godehusen, K., Martins, M., Wolff, T., \& Zimmermann, P. 2004,
  PhRvL, 93, 023002.
\newblock \url{http://dx.doi.org/10.1103/PhysRevLett.93.023002}

\bibitem[{Rudolph {et~al.}(2013)Rudolph, Bernitt, Epp, Steinbr\"ugge, Beilmann,
  Brown, Eberle, Graf, Harman, Hell, Leutenegger, M\"uller, Schlage, Wille,
  Yava\c{s}, Ullrich, \& {Crespo L\'opez-Urrutia}}]{Rudolph2013}
Rudolph, J.~K., Bernitt, S., Epp, S.~W., {et~al.} 2013, PhRvL, 111, 103002.
\newblock \url{http://dx.doi.org/10.1103/PhysRevLett.111.103002}

\bibitem[{Savin {et~al.}(2002)Savin, Kahn, Linkemann, Saghiri, Schmitt,
  Grieser, Repnow, Schwalm, Wolf, Bartsch, M{\"u}ller, Schippers, Chen,
  Badnell, Gorczyca, \& Zatsarinny}]{Savin2002c}
Savin, D.~W., Kahn, S.~M., Linkemann, J., {et~al.} 2002, ApJ, 576, 1098.
\newblock \url{http://dx.doi.org/10.1086/341810}

\bibitem[{Schippers {et~al.}(2016{\natexlab{a}})Schippers, Kilcoyne, Phaneuf,
  \& M\"uller}]{Schippers2016}
Schippers, S., Kilcoyne, A. L.~D., Phaneuf, R.~A., \& M\"uller, A.
  2016{\natexlab{a}}, ConPh, 57, 215.
\newblock \url{http://dx.doi.org/10.1080/00107514.2015.1109771}

\bibitem[{Schippers {et~al.}(2010)Schippers, Lestinsky, M{\"u}ller, Savin,
  Schmidt, \& Wolf}]{Schippers2010}
Schippers, S., Lestinsky, M., M{\"u}ller, A., {et~al.} 2010, Int. Rev. At. Mol.
  Phys., 1, 109.
\newblock
  \url{http://www.auburn.edu/academic/cosam/departments/physics/iramp/1_2/index.htm}

\bibitem[{Schippers {et~al.}(2002)Schippers, M\"uller, Ricz, Bannister, Dunn,
  Bozek, Schlachter, Hinojosa, Cisneros, Aguilar, Covington, Gharaibeh, \&
  Phaneuf}]{Schippers2002b}
Schippers, S., M\"uller, A., Ricz, S., {et~al.} 2002, PhRvL, 89, 193002.
\newblock \url{http://dx.doi.org/10.1103/PhysRevLett.89.193002}

\bibitem[{{Schippers} {et~al.}(2014){Schippers}, {Ricz}, {Buhr}, {Borovik},
  {Hellhund}, {Holste}, {Huber}, {Sch{\"a}fer}, {Schury}, {Klumpp}, {Mertens},
  {Martins}, {Flesch}, {Ulrich}, {R{\"u}hl}, {Jahnke}, {Lower}, {Metz},
  {Schmidt}, {Sch{\"o}ffler}, {Williams}, {Glaser}, {Scholz}, {Seltmann},
  {Viefhaus}, {Dorn}, {Wolf}, {Ullrich}, \& {M{\"u}ller}}]{Schippers2014}
{Schippers}, S., {Ricz}, S., {Buhr}, T., {et~al.} 2014, JPhB, 47, 115602.
\newblock \url{http://dx.doi.org/10.1088/0953-4075/47/11/115602}

\bibitem[{Schippers {et~al.}(2016{\natexlab{b}})Schippers, Beerwerth, Abrok,
  Bari, Buhr, Martins, Ricz, Viefhaus, Fritzsche, \& M\"uller}]{Schippers2016a}
Schippers, S., Beerwerth, R., Abrok, L., {et~al.} 2016{\natexlab{b}}, PhRvA,
  94, 041401.
\newblock \url{http://dx.doi.org/10.1103/PhysRevA.94.041401}

\bibitem[{Seaton {et~al.}(1994)Seaton, Yan, Mihalas, \& Pradhan}]{Seaton1994a}
Seaton, M.~J., Yan, Y., Mihalas, D., \& Pradhan, A.~K. 1994, MNRAS, 266, 805.
\newblock \url{http://dx.doi.org/10.1093/mnras/266.4.805}

\bibitem[{Simon {et~al.}(2010)Simon, {Crespo L\'opez-Urrutia}, Beilmann,
  Schwarz, Harman, Epp, Schmitt, Baumann, Behar, Bernitt, Follath, Ginzel,
  Keitel, Klawitter, Kubi\v{c}ek, M\"ackel, Mokler, Reichardt, Schwarzkopf, \&
  Ullrich}]{Simon2010a}
Simon, M.~C., {Crespo L\'opez-Urrutia}, J.~R., Beilmann, C., {et~al.} 2010,
  PhRvL, 105, 183001.
\newblock \url{https://doi.org/10.1103/PhysRevLett.105.183001}

\bibitem[{Steinbr\"ugge {et~al.}(2015)Steinbr\"ugge, Bernitt, Epp, Rudolph,
  Beilmann, Bekker, Eberle, M\"uller, Versolato, Wille, Yava\c{s}, Ullrich, \&
  {Crespo L\'opez-Urrutia}}]{Steinbruegge2015}
Steinbr\"ugge, R., Bernitt, S., Epp, S.~W., {et~al.} 2015, PhRvA, 91, 032502.
\newblock \url{http://dx.doi.org/10.1103/PhysRevA.91.032502}

\bibitem[{Stock {et~al.}(2017)Stock, Beerwerth, \& Fritzsche}]{Stock2017}
Stock, S., Beerwerth, R., \& Fritzsche, S. 2017, PhRvA, 95, 053407.
\newblock \url{https://dx.doi.org/10.1103/PhysRevA.95.053407}

\bibitem[{Thissen {et~al.}(2008)Thissen, Bizau, Blancard, Coreno, Dehon,
  Franceschi, Giuliani, Lemaire, \& Nicolas}]{Thissen2008a}
Thissen, R., Bizau, J.~M., Blancard, C., {et~al.} 2008, PhRvL, 100, 223001.
\newblock \url{https://doi.org/10.1103/PhysRevLett.100.223001}

\bibitem[{Trassl {et~al.}(1997)Trassl, Hathiramani, Broetz, Greenwood,
  McCullough, Schlapp, \& Salzborn}]{Trassl1997a}
Trassl, R., Hathiramani, P., Broetz, F., {et~al.} 1997, Phys. Scr., T73, 380.
\newblock \url{https://doi.org/10.1088/0031-8949/1997/T73/126}

\bibitem[{{Verner} \& {Yakovlev}(1995)}]{Verner1995}
{Verner}, D.~A., \& {Yakovlev}, D.~G. 1995, A\&AS, 109, 125.
\newblock \url{http://adsabs.harvard.edu/abs/1995A%26AS..109..125V}

\bibitem[{Verner {et~al.}(1993)Verner, Yakovlev, Band, \&
  Trzhaskovskaya.}]{Verner1993a}
Verner, D.~A., Yakovlev, D.~G., Band, I.~M., \& Trzhaskovskaya., M.~B. 1993,
  ADNDT, 55, 233.
\newblock \url{http://dx.doi.org/10.1006/adnd.1993.1022}

\bibitem[{Viefhaus {et~al.}(2013)Viefhaus, Scholz, Deinert, Glaser, Ilchen,
  Seltmann, Walter, \& Siewert}]{Viefhaus2013}
Viefhaus, J., Scholz, F., Deinert, S., {et~al.} 2013, NIMA, 710, 151 .
\newblock \url{http://dx.doi.org/10.1016/j.nima.2012.10.110}

\bibitem[{Wilms {et~al.}(2000)Wilms, Allen, \& McCray}]{Wilms2000}
Wilms, J., Allen, A., \& McCray, R. 2000, ApJ, 542, 914.
\newblock \url{https://doi.org/10.1086/317016}

\end{thebibliography}

\end{document}